\input amstex.tex
\documentstyle{amsppt}
\magnification\magstep1
\NoBlackBoxes

\define\gm{\bold g}
\define\hm{\bold h}
\redefine\a{\alpha}
\redefine\b{\beta}
\redefine\g{\gamma}
\redefine\s{\sigma}
\redefine\d{ d}
\redefine\R{{\Bbb R}}
\redefine\Z{{\Bbb Z}}
\redefine\C{{\Bbb C}}
\define\fii{\varphi}
\define\Fii{{\tilde\varphi}}
\define\tr{\text{tr}}

\hoffset=1cm
\voffset=1cm
\advance\vsize -1cm
\topmatter
\title  Families index theorem in supersymmetric WZW model and twisted K-theory:
The $SU(2)$ case   \endtitle

\author Jouko Mickelsson${}^1$ \\ Juha-Pekka Pellonp\"a\"a${}^2$\endauthor 
\affil ${}^1$Department of Mathematics and Statistics, University of
Helsinki; Mathematical Physics,  KTH, Stockholm \\
\it ${}^1$email: \rm  jouko.mickelsson\@helsinki.fi \\
${}^2$Department of Physics, University of Turku \\
\it ${}^2$email: \rm juhpello\@utu.fi
\endaffil 
\date revised version: June 28, 2006 \enddate
\endtopmatter

\bf Abstract. \rm The construction of twisted K-theory classes on a compact Lie group is reviewed
using the supersymmetric Wess-Zumino-Witten model on a cylinder. The Quillen superconnection
is introduced for a family of supercharges parametrized by a compact Lie group and the Chern character
is explicitly computed in the case of $SU(2).$   For large euclidean time, the character form is localized
on a D-brane. 

\document
\NoRunningHeads
\baselineskip 18pt

\vskip 0.2in

\bf 0.  Introduction \rm 

\vskip 0.2in

Gauge symmetry breaking in quantum field theory is described in  terms of
families index theory.  The Atiyah-Singer index formula gives via the Chern character
cohomology classes in the moduli space of gauge connections and of
Riemann metrics.  In particular,
the 2-form part is interpreted as the curvature of the Dirac determinant line bundle, which
gives an obstruction to gauge covariant quantization in the path integral formalism. The 
obstruction depends only on the K-theory class of the family of operators.

In the Hamiltonian quantization odd forms on the moduli space become relevant, [CMM].
The obstruction to gauge covariant quantization comes from the 3-form part of the
character. The 3-form is known as the Dixmier-Douady class and is also the (only) characteristic
class of a gerbe; this is the higher analogue of the first Chern class (in path integral quantization) 
classifying complex line bundles.  

The next step is to study families of "operators" which are only projectivly defined; that is, we have  families
of hamiltonians which are defined locally in the moduli space but which refuse to patch to a globally
defined family of operators.  The obstruction is given by the Dixmier-Douady class, an element
of integral third cohomology of the moduli space. On the overlaps of open sets the operators 
are related by a conjugation by a projective unitary transformation. This leads to the definition of
twisted K-theory, [DK], [Ro].  

In the present paper we shall first  review the basic definitions of both ordinary K-theory and twisted K-theory 
in Section 1. In Section 2 the construction of twisted (equivariant) K-theory classes on compact Lie 
groups is outlined using a supersymmetric model in $1+1$ dimensional quantum field theory.
Finally, in  Section 3 the Quillen superconnection formula is applied to the projective family 
of Fredholm operators giving a Chern character alternatively with values in Deligne cohomology
on the base or in global twisted de Rham forms, [BCMMS]. 
The use of Quillen superconnection has been proposed in general context of twisted K-theory in [Fr],
but here we  will give the details in simple terms using the supersymmetric Wess--Zumino-Witten
model. 

The Quillen superconnection can be modified (Prop.\ 1) to give a map from twisted K-theory to twisted
cohomology on the base space. However, the (nonequivariant) twisted K-theory of compact Lie
groups is all torsion [Br, Do]  and  the twisted cohomology vanishes.  Nevertheless it turns out that, at least 
in the case of the group $SU(2),$ the cocycle evaluated from the superconnection formula
contains more information than its twisted cohomology class. We show that
the de Rham form is integrally quantized as the dimension of the relevant $SU(2)$ 
representation times the basic 3-form on the group manifold. We also construct a map (not restricted 
to the case of $SU(2)$ ) from twisted K-theory to gerbes over the base, as defined in terms of local 
line bundles, modulo the twisting gerbe. In the case of $SU(2)$ this map agrees with the known identification
$K^1(SU(2),k) = \Bbb Z/k\Bbb Z,$ [Ro].

In the limit  $\to\infty$ for the scaling parameter in the superconnection, the support
of the form is localized on a 'D-brane',  a quantized conjugacy class on $SU(2),$ [GR].

\vskip 0.3in

\bf 1. Twist in K-theory by a gerbe class \rm 

\vskip 0.2in

Let $M$ be a compact manifold and $P$ a principal bundle over $M$ with
structure group $PU(H),$ the projective unitary group of a complex Hilbert space $H.$
We shall consider the case when $H$ is infinite dimensional.
The characteristic class of $P$ is represented
by an element $\Omega\in H^3(M,\Bbb Z),$ the Dixmier-Douady class. 

Choose a open cover $\{U_{\alpha}\}$ of $M$ with local transition
functions $g_{\a\b} : U_{\a\b} \to PU(H)$ of the bundle $P.$  

In the case of a good cover we can even choose lifts $\hat
g_{\a\b}:U_{\a\b} \to U(H),$ to the unitary group in the Hilbert space
$H,$ on the overlaps $U_{\a\b}= U_{\a} \cap U_{\b},$ but then we only have
$$\hat g_{\a\b}\hat g_{\b\g} \hat g_ {\g\a} =\s_{\a\b\g}\cdot \bold{1} $$
for some $\s_{\a\b\g}: U_{\a\b\g}  \to S^1$ where $U_{\a\b\g}=U_{\a} \cap U_{\b} \cap U_{\g}$ and
$\bold{1}$ is the identity operator.

Complex K-theory classes on $M$ may be viewed as homotopy classes of
maps $M\to Fred,$ to the space of Fredholm operators in an
infinite-dimensional complex Hilbert space $H.$ This defines what is
known as $K^0(M).$ The other complex K-theory group is $K^1(M)$ and
this is defined by replacing $Fred$ by $Fred_{*},$ the space of
self-adjoint Fredholm operators with both positive and negative
essential spectrum. 

The twisted K-theory classes are here defined as homotopy classes of
sections of a fiber bundle $\Cal Q$ over $M$ with model fiber equal to
either $Fred$ or $Fred_{*}.$ One sets
$$\Cal Q= P \times_{PU(H)} Fred,$$
and similarly for $Fred_{*},$ where the $PU(H)$ action on $Fred$ is
simply the conjugation by a unitary transformation $\hat g$
corresponding to $g\in PU(H).$ 

We denote by $K^{*}(M,[\Omega])$ the twisted K theory classes, the
twist given by $P.$  

Using local trivializations a section is given by a family of maps 
$\psi_{\a} : U_{\a} \to Fred$ such that 
$$\psi_{\b}(x) = \hat g_{\a\b}^{-1}(x) \psi_{a}(x)\hat g_{\a\b}(x)$$ 
on the overlaps $U_{\a\b}.$ 

\vskip 0.3in

\bf 2. Supersymmetric construction of $K(M,[\Omega])$ \rm 

\vskip 0.2in

We recall from [Mi1] the construction the   operator $Q_A$ as a sum of a 'free' supercharge $Q$ and an
interaction term $\hat A$ in (2.7) acting in $H.$ The Hilbert space $H$ is a tensor 
product of a 'fermionic' Fock space $H_f$ and a 'bosonic'  Hilbert 
space $H_b.$ Let $G$ be a connected, simply connected simple compact Lie group of dimension $N$ and $\gm$ its 
Lie algebra. The space $H_b$ carries an irreducible representation 
of the loop algebra $L\gm$ of level $k$ where the highest 
weight representations of level $k$ are classified by a finite set of $G$ 
representations (the basis of Verlinde algebra)  
on the 'vacuum sector'. 

In a Fourier basis the generators of the loop algebra are 
 $T_n^a $ where $n\in\Bbb Z$ and $a=1,..., \text{dim}\, G=N.$ 
The commutation relations are 
$$[T_n^a,T_m^b]= \lambda_{abc} T_{n+m}^{c} + \frac{k}{4}n \delta_{ab}
\delta_{n,-m},\tag2.1$$ 
where the $\lambda_{abc}$'s are  the structure constant of $\gm;$ in the case
when $\gm$ is the Lie algebra of $SU(2)$ the nonzero structure
constants are completely antisymmetric and we use the normalization 
$\lambda_{123}=\frac{1}{\sqrt{2}},$ corresponding to an orthonormal basis with respect to 
$-1$ times the Killing form. This means that in this basis the Casimir
invariant $C_2= \sum_{a,b,c} \lambda_{abc} \lambda_{acb}$ takes the
value $-N.$ 
 
In a unitary representation of the loop group we have the  hermiticity relations
 
$$(T^a_n)^*= -T^a_{-n}$$
 With this normalization of the basis, for $G=SU(2),$ $k$ is a
nonnegative integer and $2j= 0,1,2,..., k$  labels the possible irreducible representations of $SU(2)$ on the
vacuum sector. The case $k=0$ corresponds
to a trivial representation and we shall assume in the following that
$k$ is strictly positive. In general the level $k$ is quantized as
an integer $x$  times twice     the square of the length of the
longest root with respect to the dual Killing form (this unit is in
 our normalization equal to $1$ in the case $G=SU(2)$); alternatively, we
can write $k= 2x/h^{\wedge},$ where $h^{\wedge}$ is the dual Coxeter
 number of the Lie algebra $\gm.$  

The space  $H_f$ carries an irreducible representations of the
canonical anticommutation relations (CAR), 
$$\psi_n^a \psi_m^b +\psi_m^b\psi_n^a= 2\delta_{ab}\delta_{n,-m},\tag2.2$$ 
and $(\psi^a_n)^* = \psi^a_{-n}.$ The representation is fixed by the 
requirement that there is an irreducible representation of the
Clifford algebra $\{\psi_0^a\}$ in a subspace $H_{f,vac}$ such that 
$\psi_n^a v=0$ for $n<0$ and $v\in H_{f,vac}.$ 

The central extension of the loop algebra at level $2$ is
represented in $H_f$ through the operators 
$$K^a_n := -\frac14 \sum_{b,c; m\in\Bbb Z} \lambda_{abc} \psi_{n-m}^b
\psi_{m}^c, \tag2.3$$ 
which satisfy 
$$[K^a_n, K^b_m]= \lambda_{abc} K^c_{n+m} +\frac12 n \delta_{ab}\delta_{n,-m}.  
\tag2.4$$ 

We set $S^a_n= \bold1\otimes T^a_n + K^a_n \otimes\bold1.$ This gives a representation of the loop
algebra; in the case $G=SU(2)$ the  level is $k+2$ in the tensor
product $H=H_f\otimes H_b.$ In the parametrization of the level by the
integer $x$ this means that we have a level shift $x\mapsto
x'=x+h^{\wedge}.$ 
 
Next we define the supercharge operator
$$Q:= i\sum_{a,n} \left( \psi^a_n T^a_{-n} +  \frac{1}{3} \psi^a_n K^a_{-n}\right).\tag2.5$$ 
This operator satisfies $Q^2 = h,$ where $h$ is the hamiltonian of the 
supersymmetric Wess-Zumino-Witten model, 
$$h:= \underbrace{- \sum_{a,n} : T^a_n T^a_{-n} : }_{=:h_b}+ 
\frac{k+2}{2}\cdot\underbrace{\frac{1}{4} \sum_{a,n} :n \psi^a_n\psi^a_{-n}:}_{=:h_f} +\frac{N}{24}=h_b+2\tilde k h_f+\frac{N}{24},\tag2.6$$
where $\tilde k := \frac{k+2}{4}$ and 
the normal ordering $::$ means that the operators with negative
Fourier index are placed to the right of the operators with positive
index,  $:\psi^a_{-n} \psi^b_{n}:\,\, = -\psi^b_{n} \psi^a_{-n}$ if $n>0$ 
and $:AB: = AB$ otherwise. In the case of the bosonic currents $T^a_n$ 
the sign is $+$ on the right-hand-side of the equation.

Finally, the gauged supercharge operator $Q_A$ is defined as 
$$Q_A := Q + i\tilde k \sum_{a,n} \psi^a_n A^a_{-n}\tag 2.7$$ 
where the $A^a_n$'s are the Fourier components of the $\gm$-valued 
function $A$ in the basis $T^a_n$.
We denote the mapping $A\mapsto Q_A$ by $Q_\bullet$.   

The basic property of the family of self-adjoint Fredholm operators
$Q_A$ is that it is equivariant with respect to the action of the 
central extension of the loop group $LG.$ Any element $w\in LG$ is 
represented by a unitary operator $S(w)$ in $H$ but the phase of $S(w)$ 
is not uniquely determined. The equivariance property is
$$S(w^{-1}) Q_A S(w) = Q_{A^w}\tag2.8$$ 
with $A^w = w^{-1} A w +w^{-1} dw.$ For the Lie algebra we have the relations     

$$[S^a_n, Q_A] = i{\tilde k}\Big( n\psi^a_n + \sum_{b,c; m} \lambda_{abc}\psi^b_m A^c_{n-m}\Big)  
\tag2.9$$

The group $LG$ can be viewed as a subgroup of the group $PU(H)$
through the projective representation $S$ and occasionally we write $w$ instead of $S(w)$.
[In order that the embedding  is continuous in operator norm
in the Hilbert space of a positive energy representation, one should replace $LG$ by the Sobolev completion 
with respect to the weight $1/2, $ [PS].]
The space $\Cal A$ of smooth 
vector potentials on the circle is the total space for a principal 
bundle with fiber $\Omega G \subset LG,$  the group of based loops at $1.$
Since now $\Omega G \subset PU(H),$ 
$\Cal A$ may be viewed as a reduction of a $PU(H)$ principal bundle 
over $G.$ The $\Omega G$ action by conjugation on the Fredholm
operators in $H$ defines an associated fiber bundle $\Cal Q$ over 
$G$ and the family of operators $Q_A$ defines a section of this  
bundle. Thus $\{Q_A\}$ is a twisted K-theory class over $G$ where
the twist is determined by the level $k+2$ projective
representation of $LG.$  

Actually, there is additional  gauge symmetry due to constant global gauge
transformations. For this reason the construction above leads to
elements in the $G$-equivariant twisted K-theory $K^*_G(G, [\Omega]),$
where the $G$-action on $G$ is the conjugation by group elements. 
The class $[\Omega]$ is represented by the form $\frac{(k+2)}{24\pi^2}  \tr\, (g^{-1} dg)^3$ where the trace
is computed in the defining representation of $SU(2).$ 
It happens that in the case of $SU(2)$ the construction gives all generators for both equivariant and
nonequivariant twisted K-theories, but not  for other
compact Lie groups. 
 
\vskip 0.3in

\bf 3. Quillen superconnection \rm \newline

Let $Q_A$ be the supercharge associated to the vector potential $A$ on
the circle, with values in the Lie algebra $\gm.$ Recall that this
transforms as
$$ \hat g^{-1} Q_A \hat g= Q_{A^g} $$
with respect to $g\in LG\subset PU(H).$ Consider the trivial Hilbert bundle over
$\Cal A$ with fiber $H,$ the operators $Q_A$ acting in the fibers. 
Define a covariant differentiation $\nabla$ acting on the sections of 
the bundle, $\nabla:= \delta +\hat\omega$ where $\delta$ is the exterior
differentiation on $\Cal A$ and $\hat\omega$ is a connection 1-form
defined as follows. First, any vector potential on the circle can be 
uniquely written as $A=f^{-1}df$ for some smooth function $f:[0,1] \to 
G$ such that $f(0)=1$;
here we parametrize $S^1$ with parameter $y\in[0,1]$ such that any element of $S^1$ is of the form $e^{2\pi i y}$.
A tangent vector $\delta f$ at $f$ is then represented by
a function $v:[0,1] \to \gm$ such that $v(0)=0$ with periodic
derivatives at the end points and $v=f^{-1}\delta f.$  We set, [CM],
$$\omega_f(\delta f):= f^{-1}\delta f- \alpha  f^{-1} (\delta f(1) f(1)^{-1})
f\tag3.1$$ 
where $\alpha$ is a fixed smooth real valued function on $[0,1]$ such
that $\alpha(0)=0, \alpha(1)=1$ and all derivatives equal to zero 
at the end points. The point of the second term in (3.1) is that it
makes the whole expression periodic so that $\omega$ takes values in 
$L\gm.$ Then $\hat{\omega}_f(\delta f)$ is defined by the projective
representation $S$ of $L\gm$ in $H.$ 

The gauge transformation $A\mapsto A^g$ corresponds to the right translation 
$r_g(f)= fg,$ which sends $\omega_f(\delta f)$ to $g^{-1}\omega_f(\delta f) g.$ 
However, for the quantized operator $\hat \omega$ we get an additional
term. This is because of the central extension $\widehat{LG}$ which acts 
on $\hat\omega$ through the adjoint representation. One has
$$\hat g^{-1} \hat\omega \hat g= \widehat{g^{-1}\omega g} +
  \gamma(\omega,g)$$ 
with
$$\gamma(\omega,g):= \frac{k+2}{8\pi} \int_{S^1}\langle\omega, dg
g^{-1}\rangle_K.$$
The bracket $\langle\,\cdot\,,\,\cdot\,\rangle_K$ is the Killing form on $\gm.$  
But one checks that the modified 1-form
$$\hat\omega_c|_f:=\hat \omega_f - \frac{k+2}{8\pi} \int_{S^1}\langle\omega_f, f^{-1}df\rangle_K
 $$ 
transforms in a linear manner, 
$$\hat g^{-1} \hat\omega_c \hat g = \widehat{ r_g \omega_c}.\tag3.2$$ 

Here $r_g$ denotes the right action of $\Omega G$ on $\Cal A$ and the
induced right action on connection forms.
We would like to construct characteristic classes on the quotient 
space $\Cal A/\Omega G$ from classes on $\Cal A$ using the
equivariance property (3.2). 
First, we can construct a Quillen superconnecton [Qu] as the
form of mixed degree
$$D_t:=\sqrt{t} Q_\bullet + \delta + \hat\omega_c  - \frac{1}{4\sqrt{t}}  \langle\psi,F\rangle, \tag3.3$$
where $F$ is the Lie algebra valued curvature form computed from
the connection $\omega$ and $\langle\psi,F\rangle:= \sum \psi^a_n F^a_{-n}$ where $F^a_{-n}$'s are the Fourier coefficients of $F$. 
Formally, this expression is the same as the Bismut superconnection
for families of Dirac operators, [Bi]. Here $t$ is a free positive real scaling parameter.
This is introduced since in the case of Bismut superconnection one obtains
the Atiyah-Singer families index forms in the limit $t\to 0$ from the
formula (3.4) or (3.5) below.

When $\text{dim}\,G$ is even, we define 
a family of closed differential forms on $\Cal A$ from
$$\Theta^t := \tr_s \, e^{- ( \sqrt{t} Q_\bullet + \delta + \hat\omega_c -\frac{1}{4\sqrt{t}} \langle\psi,F\rangle)^2}. 
\tag3.4$$
In this case, the supertrace is defined as
$\tr_s(\cdot) = \tr\, \Gamma(\cdot).$ Here $\Gamma$ is the grading
 operator with eigenvalues $\pm 1.$ It is defined uniquely up to a phase $\pm
 1$ by the requirement that it anticommutes with each $\psi^a_n$ and
 commutes with the algebra $T^a_n.$ 
To get integral forms the
 $n$-form part of $\Theta^t$ should be multiplied by $(1/2\pi i)^{n/2}.$ 
 
 In the odd case the above
 formula has to be modified: 
$$\Theta^t := \tr^{\nu} \, e^{-(\nu \sqrt{t} Q_\bullet+ \delta + \hat\omega_c -\frac{\nu}{4\sqrt{t}}  \langle\psi,F\rangle)^2},\tag3.5$$   
where $\nu$ is an odd element, $\nu^2=1,$  anticommuting with odd differential
forms and commuting with $Q_A,$ and the trace $\tr^{\nu}$ extracts
the operator trace of the coefficients of the linear term in $\nu.$  In this case the $n$-form part
should be multiplied by $\sqrt{2i} (1/2\pi i)^{n/2}.$ 

The problem with the expressions (3.4) and (3.5) is that they cannot
pushed down to the base $G=\Cal A/\Omega G.$ The obstruction comes from
the transformation property
$$\gather  \sqrt{t} Q_{A^g} + \delta + \widehat{r_g \omega_c }  -\frac{1}{4\sqrt{t}}  \langle\psi,r_g F\rangle   \\
 = \hat g^{-1} (\sqrt{t} Q_A +\delta +
\hat\omega_c -\frac{1}{4\sqrt{t}}  \langle\psi,F\rangle)\hat g  + \hat g^* \theta, \tag3.6 \endgather $$ 
where $\theta$ is the connection 1-form on $\widehat{LG}$
corresponding to the curvature form $c$ on $LG,$ defined by the
central extension. Here $\hat g$ is a local $\widehat{LG}$ valued
function on the base $G,$ implementing a change of local section 
$G \to \Cal A.$ This additional term is the difference 
$$\hat g^* \theta= \widehat{g^{-1}\delta g} - \hat g^{-1}\delta\hat
g,$$ 
where the first term on the right comes from the transformation of the 
connection form $\hat\omega_c$ with respect to a local gauge
transformation $g.$ 
Taking the square of the transformation rule (3.6) we get 
$$(r_g D_t)^2 = \hat g^{-1} D_t^2 \hat g + g^* c.\tag3.7$$
The last term on the right arises as 
$$\delta \hat g^*\theta + (\hat g^* \theta)^2 = \hat g^* \delta \theta 
= \hat g^* c = g^* c,$$
where in the last step we have used the fact that the curvature of a
circle bundle is a globally defined 2-form $c$ on the base, and thus
does not depend on the choice of the lift $\hat g$ to $\widehat{LG}.$ 

\proclaim{Proposition 1} Let $U_{\a}$ and $U_{\b}$ be two open sets in $G$ 
with local sections $\psi_{\a},\psi_{\b}$ to the total space $\Cal A$ of the 
$\Omega G$ principal bundle $\Cal A\to G.$ Let $g_{\a\b}: U_{\a}\cap
U_{\b} \to \Omega G$ be the local gauge transformation tranforming
$\psi_{\a}$ to $\psi_{\b}.$ Then the pull-back forms $\Theta^t_{\a}$ 
and $\Theta^t_{\b}$ are related on $U_{\a}\cap U_{\b}$ as 
$$\Theta^t_{\b} =\psi^*_{\b}\Theta^t = e^{-g^*_{\a\b}c} \Theta^t_{\a}.$$ 
\endproclaim
\demo{Proof} Since the curvature is closed, $\delta c=0,$ the term
$g^* c$ on the right in (3.7) commutes with the rest and therefore can 
be taken out as a factor $\exp(-g^* c)$ in the exponential of the
square of the transformed superconnection.  \enddemo  

\bf Remark. \rm It is an immediate consequence of the Proposition 1 that the 
1-form part $\Theta^t[1]$ of $\Theta^t$ is a globally defined form on the 
base $G.$ 
We can view this as the generalization of the differential
of the families $\eta$ invariant, governing the spectral flow along 
closed loops in the parameter space; in fact, in the classical case 
of Bismut-Freed superconnection for families of Dirac operators this 
is exactly what one gets from the Quillen superconnection formula. 
We can write $\pi^{-1/2} \Theta^t[1]= h^{-1}dh/2\pi i$ with $\log h= 2\pi i \eta.$ Note that
$\eta$ is only continuous modulo integers.  Thinking of $\eta$ as the spectral asymmetry,
we normalize it by setting $\eta(A)=0$ for the vector potential $A=0,$ or on the base,
for the trivial holonomy $g=1.$

In the odd case we can relate the calculation of $\Theta^t[3]$ to the 
computation of the Deligne class in twisted K-theory, [Mi2]. 

On the overlap $U_{\a\b}$ we have from the Proposition:
$$\Theta^t_{\b}[3] =\Theta^t_{\a}[3] -g^*_{\a\b}c \wedge \Theta^t_{\a}[1].\tag3.8$$ 
 This gives
$$ \Theta^t_{\a}[3] -\Theta^t_{\b}[3]= d(\hat g^*_{\a\b}\theta \wedge
\Theta^t[1])
\equiv d\omega_{\a\b}^2.\tag3.9$$
Using $\hat g_{\a\b}\hat g_{\b\g} \hat g_{\g\a} =\s_{\a\b\g}$ we get  
$$\omega_{\a\b}^2 -\omega_{\a\g}^2 + \omega^2_{\b\g}= (\s^{-1}_{\a\b\g} 
d\s_{\a\b\g})\wedge \Theta^t[1].\tag3.10$$ 

Choose a function $h:G\to S^1$ as in the Remark,  $\pi^{-1/2} \Theta^t[1]= h^{-1}dh/2\pi i.$ 

Next the  \v{C}ech coboundary of the cochain $\{\omega^2_{\a\b}\}$ in (3.10) 
can be written as
$$d\omega^1_{\a\b\g} = d( \log(\s_{\a\b\g}) h^{-1} dh). \tag3.11$$ 
Defining $a_{\a\b\g\delta}= \log(\s_{\b\g\delta}) -   \log(\s_{\a\g\delta})
+ \log(\s_{\a\b\delta}) -\log(\s_{\a\b\g})$ we can write 
$$(\partial \omega^1)_{\a\b\g\delta} = h^{-a_{\a\b\g\delta}}dh^{a_{\a\b\gamma\delta}}\tag3.12$$
where $\partial$ denotes the \v{C}ech coboundary operator.
Thus the collection \newline
$\{\Theta^t_{\a}[3], \omega^2_{\a\b},
\omega^1_{\a\b\g}, h^{a_{\a\b\g\delta}}\}$ defines a Deligne cocycle
on the manifold $G$ with respect to the given open covering
$\{U_{\a}\}.$ 

The system of closed local forms obtained from the Chern character formula (3.5) can be modified
to a system of  \it global forms  \rm by multiplication 
$$ \tilde\Theta^t := e^{- \theta_{\a}}  \wedge \Theta^t, \tag3.13$$
where $\theta_{\a}$ is the 2-form potential, $d\theta_{\a} =\Omega$ on $U_{\a}.$  
One checks easily that now
$$(d+\Omega) \tilde\Theta^t =0. \tag3.14$$

Although the operator $d +\Omega$ squares to zero and can thus be used 
to define a cohomology theory, [BCMMS],  the Chern character should not be viewed to give a map
to this twisted cohomology theory. In fact, the twisted cohomology over complex numbers vanishes for simple
compact Lie groups. For this reason, in order to hope to get nontrivial information from
the Chern character, one should look for a refinement of the twisted cohomology.
In fact, there is another integral version of twisted cohomology proposed  in [At]. In that version one
studies the ordinary integral cohomology modulo the ideal generated by the Dixmier-Douady class
$\Omega.$ At least in the case of $SU(2)$ it is an experimental fact that the twisted (nonequivariant) 
K-theory as an abelian
group is isomorphic to the twisted cohomology in this latter sense.
There is a similar result for
other compact Lie groups, but the 3-cohomology class used to define the twisting in cohomology 
is in general not the original twisting gerbe class; both are integral multiples of a basic 3-form, but the coefficients
differ, except for the case of $SU(2),$ [Br, Do].   

One can explicitly see why the integral cohomology mod $\Omega$ is relevant for twisted K-theory
by the following construction in the odd case.  First we replace the space $Fred_*$ by the homotopy equivalent
space $\Cal{U}_1$ consisting of $\bold{1}\,+\,$trace-class unitaries on $H.$ An ordinary K-theory class on $X$ is a homotopy
class of maps $m:X \to \Cal{U}_1.$ In this representation the Chern character  
defines a sequence of cohomology classes on $X$ by pulling back the generators
$\tr\, (g^{-1}dg)^{2n+1}$ of the cohomology of $\Cal{U}_1.$  In the twisted case we have only maps $m_\a$ on open sets $U_{\a}$ which are related 
by $m_{\b} = g_{\a\b}^{-1} m_{\a} g_{\a\b}$ on overlaps.

In the case of $G=SU(2)=S^3$ we need only two open sets $U_{\pm},$ the slightly extended upper and lower hemispheres, and a map 
$g_{-+}$ on the overlap $U_{-+}=U_{-}\cap U_{+}$ to the group $PU(H),$ of degree $k.$ 
If now $m_-\equiv\bold{1}$ identically and 
the set of points $x$ with $m_+(x)\neq\bold1$ (the support of $m_+$) 
is concentrated around the North pole, then the pair $m_{\pm}$ is related by the
conjugation by $g=g_{-+}$ at the equator and at the same time it defines an ordinary K-theory class since the functions
patch to a globally defined function on $S^3.$ Let us assume that the winding number of $m: S^3 \to \Cal{U}_1$ is $k.$ 
Next form a continuous path $m_{\pm}(t)$ of representatives of twisted K-theory classes starting from $m(1)=m$ and ending at 
the trivial class represented by the constant function $m(0)=\bold{1}.$ Let $\rho$ be a smooth function on $S^3$ which is equal to
$1$ on the overlap $U_{-+}$ and zero in small open neighborhoods $V_{\pm}$  of the poles. 
We can also extend the domain of definition of $g(x)$ 
to a larger set  $U_+\setminus V_+.$ 
For $0\leq t \leq 1$ define 
$$[m_-(t)](x):= e^{2\pi it \rho(x)P_0}
 \text{ and } [m_+(t)](x) := e^{  2\pi  it \rho(x)  P(x) }$$
where $P(x)= g^{-1}(x) P_0 g(x),$ with $P_0$ is a fixed rank one projection, $x\in U_{-+}.$ 
These are smooth functions on $U_{\pm}$ respectively and are related by the conjugation by $g$ on the overlap.
But for $t=0$ both are equal to the identity $\bold{1}\in \Cal{U}_1.$ On the other hand, at $t=1$ the integral 
$$\frac{1}{24 \pi^2} \int_{S^3} \tr\, (m^{-1} dm)^3$$
is easily computed to give the value $k.$  This paradox is explained by the fact that for the intermediate values
$0 < t < 1	$ the functions $m_{\pm}$ do not patch up to a global function on $S^3.$  Thus we have a homotopy joining
a pair of (trivial) twisted K-theory classes corresponding to the pair of third cohomology classes $0, \Omega$ 
computed from the Chern character. This confirms the claim, at least in the case of $X=S^3,$ that the values 
of the Chern character should be projected to the quotient $H^*(X,\Bbb Z)/\Bbb Z \Omega.$ 
\vskip 0.3in

\bf 4. Quillen superconnection in the case of $SU(2)$\rm \newline

In this section, we consider the case $G=SU(2)$ and calculate a pull-back of $\Theta^t=\tr^\sigma e^{-D_t^2},$ with respect to a local section 
over $U_+ = SU(2)\setminus \{-1\},$ in the limit $t\to\infty$.

The basic idea is that in this case one can naturally associate to the twisted K-theory class a K-theory class, defined modulo the twisting gerbe.
This method was used in [Mi2] to calculate a characteristic class for twisted K theory on 3-manifolds.
Actually, on the level of 3-cohomology that method can be generalized to arbitrary simply connected compact Lie groups as follows.

Let $\{U_{\a}\}$ be an open cover of $G$ which trivializes the bundle $\Cal A \to G.$ Let $\psi_{\a} :U_{\a} \to \Cal A$ be local sections. 
We may take $\{U_{\a}\}$ as the open set of holonomies such that the real number $\a$ is not in the spectrum of $Q_{A}$ (for any $A$ in 
the fibers over $U_{\a}$). Let $L'_{\a\b}(A)$ be be the top exterior power  of the finite dimensional spectral subspace $E_{\a\b}:\a < Q_A < \b$ 
and denote 
$$L_{\a\b} = \psi^*_{\a} L'_{\a\b} \otimes (h^*_{\a\b} c)^{-n_{\b}}$$ 
where $c$ is the level $\tilde k$ central extension cocycle of the loop group and $h_{\a\b}$ is the transition function, 
$\psi_{\b} =\psi_{\a} h_{\a\b}.$ The integers $n_{\b}$ will be determined below. We require that 
$$L_{\a\b}\otimes L_{\b\g} \otimes L_{\g\a}= \Bbb C, $$
canonically isomorphic to the trivial line bundle. First note that $\psi_{\b}^* L'_{\b\g} = \psi_{\a}^* L'_{\b\g} \otimes (h_{\a\b}^* c)^{n_{\b\g}}$ 
by the projective action of the loop group on the spectral subspaces, where $n_{\b\g}= \text{dim } E_{\b\g}$ for $\b <\g$ and $n_{\g\b}=-n_{\b\g}.$ 
Therefore 
$$\align L_{\a\b} & \otimes L_{\b\g}\otimes L_{\g\a} = \psi^*_{\a} L'_{\a\b}\otimes \left(\psi^*_{\a} L'_{\b\g} \otimes (h^*_{\a\b} c)^{n_{\b\g}}
\right) \\
& \otimes \left(\psi^*_{\a} L'_{\g\a}\otimes (h^*_{\a\g} c)^{n_{\g\a}}\right) \otimes(h^*_{\a\b} c)^{-n_{\b}}\otimes (h^*_{\b\g} c)^{-n_{\g}} 
\otimes (h^*_{\g\a} c)^{-n_{\a}} \\ 
& = \psi_\a^*(L'_{\a\b} \otimes L'_{\b\g} \otimes L'_{\g\a} ) \otimes (h^*_{\a\b} c)^{n_{\b\g}-n_{\b}} \otimes (h^*_{\b\g}c)^{-n_{\g}} 
\otimes (h^*_{\g\a}c)^{-n_{\g\a}-n_{\a}}.\endalign$$
The line bundles $L'_{\a\b}$ form a cocycle as follows from $E_{\a\g} = E_{\a\b} \oplus E_{\b\g}$ and therefore the nontrivial part is a product of pull-backs of $c.$ 
Using the group property of the central extension of the loop  group we have $h^*_{\a\g} c = h^*_{\a\b}c \otimes h^*_{\b\g} c$ and collecting the factors the 
tensor product above becomes
$$ (h^*_{\a\b} c)^{n_{\b\a} - n_{\b} + n_{\a}}\otimes (h^*_{\b\g}c)^{n_{\g\a} -n_{\g} +n_{\a}},$$
where we have taken into account $n_{\b\g}+n_{\g\a}=n_{\b\a}$ in the first factor,
and this becomes trivial provided we can choose the locally constant functions $n_{\a}$ such that $n_{\a\b} = n_{\a} - n_{\b}$
on triple intersections. But since we assumed that the base $G$ is simply connected, $H^1(G)=0$ and the solution exists. 

The solution is not uniquely defined: it is defined modulo adding to each $n_{\a}$ a constant. This corresponds to modifying the gerbe defined 
by the line bundles $L_{\a\b}$ by a power of the gerbe defined by the local line bundles $h^*_{\a\b}c.$ But these latter line bundles correspond 
to the Dixmier-Douady class $\Omega.$ Thus we get a map from twisted K theory $K^1(G, \Omega)$ to $H^3(G,\Bbb Z)/\Bbb Z\Omega.$ This map needs 
be neither injective nor surjective. However, in the case of $G=SU(2)$ these groups are known to be equal.   

We now explain why the Quillen superconnection formalism has to give the same result. The calculation of the characteristic class in
[Mi2] was based on the following observation. As a classifying space for self-adjoint Fredholm operators, with both positive and negative 
essential spectrum, one can also use the Lie group $G_1$ consisting of unitary operators which differ from the unit by a trace class 
operator. The 3-cohomology of $G_1$ is generated by $\tr\, (g^{-1}dg)^3.$ In general, the $K^1$ class of a map $\phi: X \to G_1$ is mapped 
to cohomology as the odd Chern character given by the de Rham forms $\phi^* \tr\, (g^{-1}dg)^{2i+1}.$ 
In the twisted case we do not have a map from $X$ to $G_1$ but a family of local maps which are related on intersections of open sets 
by conjugation by $PU(H)$ valued transition functions. But in the case $X=SU(2)$ one can deform the twisted map such that it becomes 
a globally well-defined map $SU(2) \to G_1 $ and one gets a characteristic class $\phi^* \tr\,(g^{-1}dg)^3$ in twisted K theory.
However, as explained in the end of previous Section, this class is defined only modulo the Dixmier-Douady class of the twisting bundle. 

Going back to the $G=SU(2)$ case, in [Mi2] the map from twisted K-theory to untwisted K theory was explicitly performed by using the fact 
that the $\Omega G$ bundle over $G$ is trivialized over a pair of open sets, $S^3_+ = S^3\setminus \{N\}$ and $D^3_N$  
where $S,N$ are the 'South' and 'North' poles on the 3-sphere and $D^3_N$ is a small disk around the North pole.  The transition functions 
are defined on a thickened 2-sphere close to $N.$  After a deformation, the map $g_+: S^3_+ \to G_1$ becomes constant equal to 
the unit element in $G_1$ and thus defines a global $K^1$ class on $S^3$ by pull-back from $G_1.$ 
Applying the Quillen superconnection to this class (using instead of $G_1$ the space $Fred_*$ as a classifying space) must give the same 
cohomology class. However, we can do better than that: 

Using the global $SU(2)$ gauge invariance of the construction of Fredholm operators, we know that the differential forms obtained
from the Quillen superconnection must be invariant under conjugation by $SU(2).$ The cohomology class does not depend on the scaling 
parameter $t.$ On the other hand, in the limit $t\to\infty$ the forms are supported only on the subset of parameters $A$ such that 
the operator $Q_A$ has zero eigenvalue. The zeros of $Q_A$ are easily computed, [FHT2]. The elements of a simple compact Lie group $G$  
can be parametrized  as $g=\exp(2\pi a),$ where $a$ is in the Lie algebra
of $G.$ The constant vector potential $A\equiv a$ on the circle $S^1$ has the holonomy $g.$ 
Since the construction of the operators $Q_A$  is equivariant with respect to (constant) gauge 
transformations we may assume that $a$ is in the Cartan subalgebra $\bold h$ of $G.$ 
Actually, performing an additional gauge transformation one can require that 
$$(a+  \overset{\vee}\to{d} , \alpha_i) \leq 0 $$
where $\alpha_i$'s are the simple roots of the affine Lie algebra based on $G,$   $\vee$  means the duality transformation $\bold h\to \bold h^*$ 
defined by the Killing form, and $(\cdot,\cdot)$ is the inner product in $\bold h^*$ coming from the 
Killing form. Here $d=-i \frac{d}{d\theta}$ is the derivation of the loop algebra. 
Then the only zeros of $Q_A$ are in the weight subspace $\lambda + \rho$ where $\lambda$ is
the highest weight of the $G$ representation on the bosonic vacuum sector, $\rho$ is half the sum
of the positive roots (which is also a highest weight on the fermionic vacuum sector). The value of $a$ 
corresponding to the nonzero kernel of $Q_A$ is given by
$$ \tilde k \overset{\vee}\to{a} =  -\lambda -\rho. $$
In particular, for $G=SU(2)$ we obtain $\overset{\vee}\to{a} = -2j/\tilde k$ where $2j=0,1,2,\dots$ and in our normalization 
$\tilde k= (k+2)/4.$ 

Let $\fii\in [0,2\pi[$ and $\bold n\in S^2$.
 Define a local section $\psi: SU(2)\setminus \{N\} \to \Cal A$ by 
$ \psi(\fii, \bold n) = -\frac{i}{2} \fii \bold n\cdot\bold\sigma$ where $\bold\sigma=(\sigma_1,\sigma_2,\sigma_3)$ and $\sigma_a$'s are the Pauli matrices.
At the boundary $\fii=2\pi$ all the constant 
vector potentials $\psi(2\pi,\bold n)$ have holonomy $-1$ around the circle. 

At $\fii=2\pi$ the operators $Q_{\psi(\fii,\bold n)}$ are invertible and for this reason at the limit 
$t\to\infty$ the exponential of the Quillen superconnection vanishes: Thus in this limit the forms obtained 
from (3.5) become globally well-defined on $SU(2)$ despite the fact that $\psi: SU(2) \to \Cal A$ is 
discontinuous at $\fii=2\pi.$ 

To conclude, since the 3-form part of $\Theta^t$ invariant under conjugation by $SU(2)$ and, 
in the limit $t\to\infty$, $\psi^*\Theta^t[3]$ becomes concentrated on the 2-sphere, the form must be proportional to the area 2-form of the sphere. The proportionality constant is uniquely determined by the cohomology class. This justifies the result of Theorem.

Let $K$ be a compact (Hausdorff) space, and let $C(K)$ be a Banach space of continuous functions $u:\,K\to\C$ equipped with the sup norm.
By the Riesz representation theorem, the topological dual $M(K)$ of $C(K)$ can be considered as a set of regular complex Borel measures on $K$.
Equip $M(K)$ with the weak-star topology.
In this topology, we say that a net $\{\mu_i\}_{i\in\Cal I}\subseteq M(K)$, where $\Cal I$ is a directed set, converges to a point
$\mu\in M(K)$ if $\lim_{i\in\Cal I}\int u\d\mu_i=\int u\d\mu$ for any $u\in C(K)$; we denote $\mu=\mathop{\hbox{w$^*$-lim}}_{i\in\Cal I}\mu_i$.

Let ${\bold{area}}(S^2)\in M(S^2)$ denote the measure defined by the area 2-form of $S^2$, 
$\delta_a\in M\big([0,2\pi]\big)$ the Dirac measure concentrated on $a\in \, ]0,2\pi[$, and
$$
\fii_j^k:=2\pi\frac{2j+1}{4\tilde k}=2\pi\frac{2j+1}{k+2}\in \, ]0,2\pi[.
$$

\proclaim{Theorem} 
When $s^*\Theta^t[3]$ is interpreted as a measure on $[0,2\pi]\times S^2$
$$
\mathop{\hbox{\rm w$^*$-lim}}_{t\to\infty}\psi^*\Theta^t[3]=-\sqrt{\pi}i\left(j+\frac{1}{2}\right)\delta_{\fii_j^k}\otimes{\bold{area}}(S^2).
$$

\endproclaim
\noindent
To get an integral form, we multiply $\Theta^t[3]$ by $(\sqrt{\pi}i)^{-1}(2\pi)^{-1}$. 

For completeness, next we represent the sketch of the direct but lengthy proof of Theorem  taken from our previous (unpublished) version of the article, [MP].

\vskip 0.1in

\noindent
\bf Proof. \rm Since
$$\align
D_t^2=& t Q_\bullet^2 +\sqrt{t}\nu\underbrace{\Big(-\delta Q_\bullet+[Q_\bullet,\hat\omega_c]\Big)}_{=:B_1\,\hbox{(1-form)}}+
\underbrace{\Big(\delta\hat\omega_c+\hat\omega_c^2-\frac{1}{4}\{Q_\bullet,\langle\psi,F\rangle\}\Big)}_{=:B_2\,\hbox{(2-form)}}
\\
&-\frac{\nu}{4\sqrt{t}}\underbrace{\Big(-\delta\langle\psi,F\rangle+[\langle\psi,F\rangle,\hat\omega_c]\Big)}_{=:B_3\,\hbox{(3-form)}} 
=t\big(Q_\bullet^2-E_t\big) \tag4.2
\endalign $$
where
$$
E_t:=-\frac{1}{t}\left(\sqrt{t}\nu B_1+B_2-\frac{\nu}{4\sqrt{t}} B_3\right), \tag4.3
$$
using the perturbation series expansion 
$$
\tr^\nu e^{-t(Q^2_\bullet-E_t)}=\tr^\nu e^{-t Q^2_\bullet}+\sum_{n=1}^\infty t^n\int_{\Delta_n}\tr^\nu\left(e^{-t s_1Q^2_\bullet}E_te^{-t s_2Q^2_\bullet}
E_t\cdots e^{-t s_{n+1}Q^2_\bullet}\right)\d s_1\cdots\d s_n,
$$
where $\Delta_n$ is the standard $n$-simplex,
one gets
$$\align
\Theta^t=&\tr^\nu e^{-t Q^2_\bullet}
+t\int_{\Delta_1}\tr^\nu\left(e^{-t s_1Q^2_\bullet}E_te^{-t s_2Q^2_\bullet}\right)\d s_1
\\
&+t^2\int_{\Delta_2}\tr^\nu\left(e^{-t s_1Q^2_\bullet}E_te^{-t s_2Q^2_\bullet}E_t e^{-t s_3 Q^2_\bullet}\right)\d s_1\d s_2 
\\
&+t^3\int_{\Delta_3}\tr^\nu\left(e^{-t s_1Q^2_\bullet}E_te^{-t s_2Q^2_\bullet}E_t e^{-t s_3 Q^2_\bullet}E_te^{-t s_4 Q^2_\bullet}\right)\d s_1\d s_2\d s_3. 
\\
\endalign $$
The three form part of the above form is
$$\align
\Theta^t[3]=&t\sqrt{t}\int_{\Delta_3}\tr\left(e^{-t s_1Q^2_\bullet}B_1 e^{-t s_2Q^2_\bullet}B_1 e^{-t s_3 Q^2_\bullet}B_1 e^{-t s_4 Q^2_\bullet}\right)\d s_1\d s_2\d s_3\\
&+\sqrt{t}\int_{\Delta_2}\tr\left(e^{-t s_1Q^2_\bullet}B_1e^{-t s_2Q^2_\bullet}B_2 e^{-t s_3 Q^2_\bullet}
+e^{-t s_1Q^2_\bullet}B_2e^{-t s_2Q^2_\bullet}B_1 e^{-t s_3 Q^2_\bullet}\right)\d s_1\d s_2\\
&\frac{1}{4\sqrt{t}}\int_{\Delta_1}\tr\left(e^{-t s_1Q^2_\bullet} B_3 e^{-t s_2Q^2_\bullet}\right)\d s_1.
\endalign $$
After long simplification (see, [MP]) one gets
$$\align
\psi^*\Theta^t[3]_{\fii,\bold n}=&
-\frac{i\tilde k^2}{\pi^2\sqrt{2}}\sqrt{t}e^{-t[2{\tilde k}^2\Fii^2-(2j+1)\tilde k\Fii +j(j+1)/2+1/8]}\times\\
&\times\left\{
\big[\fii_j^k+(\fii-\fii_j^k)F_\alpha(\fii)\big]
\,\d\fii\wedge{\bold{area}}(S^2)|_{\bold n}
+{\Cal O}(t^{-1})\right\}
\endalign $$
where $\tilde\fii=\fii/(2\pi)$,
$F_\alpha:\,[0,2\pi)\to\R$ is continuous (depends on $\alpha$) and
 $${\bold{area}}(S^2)|_{\bold n}:=n_1\d n_2\wedge\d n_3+n_2\d n_3\wedge\d n_1+n_3\d n_1\wedge\d n_2=\sin\theta\,\d\theta\wedge\d\phi$$ is the area 2-form of $S^2$ at $\bold n$ (where $(\theta,\phi)$ are the spherical coordinates).

Using the formal notation $\delta(\fii-a)$ for the Dirac measure $\delta_a$
concentrated on $a\in(0,2\pi)$ 
one gets
$$
\delta(\fii-a)=\mathop{\hbox{w$^*$-lim}}_{r\to\infty}\sqrt{\frac{r}{\pi}}e^{-r(\fii-a)^2}
$$
and for any $p\ge 1$,
$$
\mathop{\hbox{w$^*$-lim}}_{r\to\infty}\frac{1}{r^p}\sqrt{r}\,e^{-r\fii^2}=0.
$$
Since $2{\tilde k}^2\Fii^2-(2j+1)\tilde k\Fii +j(j+1)/2+1/8=\left(\frac{\tilde k}{\sqrt{2}\pi}\right)^2(\fii-\fii_j^k)^2$, by putting the parameter $r$ to $\left(\frac{\tilde k}{\sqrt{2}\pi}\right)^2t$ in the above equations, we may conclude that,
when $\psi^*\Theta^t[3]$ is interpreted as a measure on $[0,2\pi]\times S^2$,
$$
\mathop{\hbox{\rm w$^*$-lim}}_{t\to\infty}\psi^*\Theta^t[3]=-\sqrt{\pi}i\left(j+\frac{1}{2}\right)\delta_{\fii_j^k}\otimes{\bold{area}}(S^2)
$$
and Theorem  follows.


\vskip 0.3in

\bf References \rm 

\vskip 0.2in
[At] M. Atiyah: K theory past and present.  math.KT/0012213.
Sitzungsberichte der Berliner  Mathematischen Gesellschaft,  411--417, Berliner Math.\ Gesellschaft, Berlin, 2001.

[AS] M. Atiyah and G. Segal:  Twisted K-theory. math.KT/0407054
 
[Bi] J.-M. Bismut: Localization formulas, superconnections, and the index
 theorem for families.   Comm. Math. Phys.  \bf 103, \rm  no. 1, 127--166 (1986) 

[BCMMS] P. Bouwknegt, A. Carey,  V. Mathai,  M. K. Murray, and D. Stevenson: Twisted K-theory and K-theory of
bundle gerbes.  hep-th/0106194. Commun. Math. Phys.  \bf 228, \rm  17-49 (2002)

[Br] Volker Braun: Twisted K theory of Lie groups. 
hep-th/0305178. \bf JHEP \rm 0403, 029 (2004)

[CM] A. L. Carey and J. Mickelsson: The universal gerbe, Dixmier-Douady class, and gauge theory.
    Lett. Math. Phys. \bf 59, \rm   47-60 (2002)

[CMM] A. L. Carey, J. Mickelsson, and M.K. Murray: Index theory, gerbes,
and hamiltonian quantization hep-th/9511151
Commun. Math. Phys. \bf 183, \rm 707-722 (1997)

[DK]  P. Donovan and M. Karoubi: Graded Brauer groups and K-theory with local coefficients.  Inst. Hautes Études Sci.
Publ. Math. No. \bf 38, \rm  5-25 (1970)

[Do] Christopher L. Douglas:  On the twisted K-homology of simple Lie groups.  math.AT/0402082

[FdV]  H. Freudenthal and H. de Vries: \it Linear Lie Groups. \rm Pure Appl. Math. \bf 35, \rm Academic Press, 
New York (1969)

[Fr]  D. Freed: Twisted K-theory and loop groups. math.AT/0206237. Publ. in the proceedings of
ICM2002, Beijing.

[FHT] D. Freed, M. Hopkins, and C. Teleman: Twisted  equivariant K-theory with complex coefficients.
math.AT/0206257. Twisted K-theory and loop group representations.  math.AT/0312155

[GR] Krzysztof Gawedzki and  Nuno Reis: WZW branes and gerbes. hep-th/0205233. Rev. Math. Phys. \bf 14, \rm   1281-1334 (2002)

[Mi1]  J. Mickelsson: Gerbes, (twisted) K theory, and the supersymmetric WZW model. hep-th/0206139.
Publ. in  \it Infinite Dimensional Groups and Manifolds,   \rm  ed. by T. Wurzbacher. IRMA Lectures in
Mathematics and Theoretical Physics \bf 5, \rm Walter de Gruyter, Berlin (2004)

[Mi2]  J. Mickelsson: Twisted K theory invariants.  math.AT/0401130. Lett. in Math. Phys. \bf 71, \rm  109-121 (2005)

[MP] J. Mickelsson and J.-P. Pellonp\"a\"a: Families index theorem in supersymmetric WZW model and twisted K-theory: The $SU(2)$ case. hep-th/0509064, version 1

[PS] A. Pressley and G. Segal: \it Loop Groups. \rm Clarendon Press, Oxford (1986)

[Qu] D. Quillen: Superconnections and the Chern character.  Topology
\bf 24, \rm  no. 1, 89--95 (1985)

[Ro] J. Rosenberg:  Continuous-trace algebras from the bundle theoretic point of view.  J. Austral. Math. Soc. Ser. \bf A  47 \rm  no. 3, 368--381 (1989) 

\enddocument

\vskip 0.3in

\bf References \rm 

\vskip 0.2in
[At] M. Atiyah: K theory past and present.  math.KT/0012213.
Sitzungsberichte der Berliner  Mathematischen Gesellschaft,  411--417, Berliner Math. Gesellschaft, Berlin, 2001.

[AS] M. Atiyah and G. Segal:  Twisted K-theory. math.KT/0407054
 
[Bi] J.-M. Bismut: Localization formulas, superconnections, and the index
 theorem for families.   Comm. Math. Phys.  \bf 103, \rm  no. 1, 127--166 (1986) 

[BCMMS] P. Bouwknegt, A. Carey,  V. Mathai,  M. K. Murray, and D. Stevenson: Twisted K-theory and K-theory of
bundle gerbes.  0106194. Commun. Math. Phys.  \bf 228, \rm  17-49 (2002)

[CMM] A. Carey, J. Mickelsson, and M.K. Murray: Index theory, gerbes,
and hamiltonian quantization hep-th/9511151
Commun.Math.Phys.\bf 183, \rm 707-722 (1997)

[Do] Christopher L. Douglas:  On the twisted K-homology of simple Lie groups.  math.AT/0402082

[FdV]  H. Freudenthal and H. de Vries: \it Linear Lie Groups. \rm Pure Appl. Math. \bf 35, \rm Academic Press, 
New York (1969)

[Fr]  D. Freed: Twisted K-theory and loop groups. math.AT/0206237. Publ. in the proceedings of
ICM2002, Beijing.

[FHT] D. Freed, M. Hopkins, and C. Teleman: Twisted  equivariant K-theory with complex coefficients.
math.AT/0206257. Twisted K-theory and loop group representations.  math.AT/0312155

[Mi1]  J. Mickelsson: Gerbes, (twisted) K theory, and the supersymmetric WZW model. hep-th/0206139.
Publ. in  \it Infinite Dimensional Groups and Manifolds,   \rm  ed. by T. Wurzbacher. IRMA Lectures in
Mathematics and Theoretical Physics \bf 5, \rm Walter de Gruyter \ , Berlin (2004)

[Mi2]  J. Mickelsson: Twisted K theory invariants.  math.AT/0401130. Lett. in Math. Phys. \bf 71, \rm  109-121 (2005)

[Qu] D. Quillen: Superconnections and the Chern character.  Topology
\bf 24, \rm  no. 1, 89--95 (1985)

\enddocument

\vskip 0.3in
\bf Appendix: Zeros of $Q_A$ \rm 

\vskip 0.2in

The zero modes of the operator $Q_A$ are the same as its square $Q_A^2$ but the latter is easier to handle.
First we gauge transform the potential $A$ to a constant potential in the Cartan subalgebra $\hm$ and denote by
$\mu\in \hm^*$ its dual with respect to the Killing form. Then 
$$ Q_A^2 = h + \frac{N}{24} + |\tilde k \mu|^2 + 2\tilde k \mu_i h^{i}   \tag{A1}.$$

Acting with $h$ on the highest weight vectors, corresponding to the weight $\lambda +\rho$  in $H_b\otimes H_f,$
gives the eigenvalue  $|\lambda +\rho|^2 -|\rho|^2.$ Here $\rho$ is half the sum of positive roots of $\gm.$ 
In our normalization of the inner product,
$|\rho|^2 = N/24,$ [FdV], and therefore the eigenvalue of  $Q_A^2$ on the highest weight vector is 
$$ Q_A^2 (\lambda) = |\lambda +\rho|^2 + \tilde k^2|\mu|^2 + 2\tilde k (\mu, \lambda +\rho).\tag{A2}$$

When acting on a general vector of weight $\lambda'$ in the representation space the eigenvalue of $h$ is shifted by
$\tilde k(\lambda' -\lambda - \rho)(2d)$ and so the eigenvalue of $Q_A^2$ becomes
$$ Q_A^2(\lambda') =  |\lambda +\rho|^2 + \tilde k^2 |\mu|^2+ 2\tilde k (\mu, \lambda') + 2\tilde (\lambda' -\lambda -\rho)(d).
\tag{A3}$$
We can rewrite this expression as 
$$Q_A^2 (\lambda') = |\lambda +\rho + \tilde k \mu|^2 -+2\tilde k (\lambda' -\lambda -\rho, \mu + \text{\v{$d$}} ), \tag{A4}$$
where \v{$d$} is the dual root of $d.$ 
After action by a Weyl group element of the affine Lie algebra (implemented here by a gauge transformation) we may assume 
that $\mu + \text{\v{$d$}}$ is in the  Weyl  chamber defined by
$$ (\mu + \text{\v{$d$}}, \alpha_i) \leq 0,$$ 
where $\alpha_i$ with $i=0,1,2,\dots \ell$ are the simple roots. 
Since the difference $\lambda +\rho -\lambda'$ is a sum of simple roots, we conclude that the second term in (A4) is nonnegative.
Thus in order that the eigenvalue $Q_A^2(\lambda')=0$ the first term has to vanish. This implies
$$\tilde k \mu = - \lambda -\rho. \tag{A5}$$

From the last equation follows that all the coordinates $(\mu,\alpha_i)$ are strictly negative , by $(\rho,\alpha_i)=1,$ and so 
the second term in (A4) vanishes if and only if $\lambda' -\lambda -\rho = 0.$ We conclude that the only zeros of $Q_A$ are in
the vacuum $\lambda'=\lambda + \rho$ when $\mu +\text{\v{$d$}}$ is in the negative of the fundamental Weyl chamber  and then 
$\mu$ is given by (A5).

Denote the superconnection $\sqrt{t}\sigma Q_\bullet + \delta + \hat\omega_c  - \frac{\sigma}{4\sqrt{t}}\langle\psi,F\rangle$ by $D_{1,t}$
and define a new superconnection $D_{2,t}:=\sqrt{t}\sigma Q_\bullet + \delta$.
Choose a one parameter family of superconnections joining them: redefine
$$D_t:=u D_{1,t}+(1-u) D_{2,t}=\sqrt{t}\sigma Q_\bullet + \delta + u\hat\omega_c  - \frac{u\sigma}{4\sqrt{t}}\langle\psi,F\rangle. \tag4.1$$
where $u\in[0,1]$ is here considered as a constant, i.e., $\delta u=0$.
Thus,
$$\align
D_t^2=& t Q_\bullet^2 +\sqrt{t}\sigma\underbrace{\Big(-\delta Q_\bullet+u[Q_\bullet,\hat\omega_c]\Big)}_{=:B_1\,\hbox{(1-form)}}+
\underbrace{\Big(u\delta\hat\omega_c+u^2\hat\omega_c^2-\frac{u}{4}\{Q_\bullet,\langle\psi,F\rangle\}\Big)}_{=:B_2\,\hbox{(2-form)}}
\\
&-\frac{\sigma}{4\sqrt{t}}\underbrace{\Big(-u\delta\langle\psi,F\rangle+u^2[\langle\psi,F\rangle,\hat\omega_c]\Big)}_{=:B_3\,\hbox{(3-form)}} 
=t\big(Q_\bullet^2-E_t\big) \tag4.2
\endalign $$
where
$$
E_t:=-\frac{1}{t}\left(\sqrt{t}\sigma B_1+B_2-\frac{\sigma}{4\sqrt{t}} B_3\right). \tag4.3
$$

Using the perturbation series expansion 
$$
\tr^\sigma e^{-t(Q^2_\bullet-E_t)}=\tr^\sigma e^{-t Q^2_\bullet}+\sum_{n=1}^\infty t^n\int_{\Delta_n}\tr^\sigma\left(e^{-t s_1Q^2_\bullet}E_te^{-t s_2Q^2_\bullet}
E_t\cdots e^{-t s_{n+1}Q^2_\bullet}\right)\d s_1\cdots\d s_n,
$$
where $\Delta_n$ is the standard $n$-simplex
$$
\left\{(s_1,s_2,\ldots,s_{n+1})\in\R^{n+1}\,\bigg|\,0\leq s_a\leq 1,\;\sum_{a=1}^{n+1} s_a=1\right\},
$$
one gets
$$\align
\Theta^t=&\tr^\sigma e^{-t Q^2_\bullet}
+t\int_{\Delta_1}\tr^\sigma\left(e^{-t s_1Q^2_\bullet}E_te^{-t s_2Q^2_\bullet}\right)\d s_1
\\
&+t^2\int_{\Delta_2}\tr^\sigma\left(e^{-t s_1Q^2_\bullet}E_te^{-t s_2Q^2_\bullet}E_t e^{-t s_3 Q^2_\bullet}\right)\d s_1\d s_2 
\\
&+t^3\int_{\Delta_3}\tr^\sigma\left(e^{-t s_1Q^2_\bullet}E_te^{-t s_2Q^2_\bullet}E_t e^{-t s_3 Q^2_\bullet}E_te^{-t s_4 Q^2_\bullet}\right)\d s_1\d s_2\d s_3 
\\
=&\Theta^t[1]+\Theta^t[3]
\endalign $$
where $\Theta^t[3]:=\Theta_1^t[3]+\Theta_2^t[3]+\Theta_3^t[3]$ and
$$\align
\Theta^t[1]:=&{-\sqrt{t}\int_{\Delta_1}\tr\left(e^{-t s_1Q^2_\bullet}B_1e^{-t s_2Q^2_\bullet}\right)\d s_1},\\
\Theta_1^t[3]:=&t\sqrt{t}\int_{\Delta_3}\tr\left(e^{-t s_1Q^2_\bullet}B_1 e^{-t s_2Q^2_\bullet}B_1 e^{-t s_3 Q^2_\bullet}B_1 e^{-t s_4 Q^2_\bullet}\right)\d s_1\d s_2\d s_3,\\
\Theta_2^t[3]:=&\sqrt{t}\int_{\Delta_2}\tr\left(e^{-t s_1Q^2_\bullet}B_1e^{-t s_2Q^2_\bullet}B_2 e^{-t s_3 Q^2_\bullet}
+e^{-t s_1Q^2_\bullet}B_2e^{-t s_2Q^2_\bullet}B_1 e^{-t s_3 Q^2_\bullet}\right)\d s_1\d s_2,\\
\Theta_3^t[3]:=&\frac{1}{4\sqrt{t}}\int_{\Delta_1}\tr\left(e^{-t s_1Q^2_\bullet} B_3 e^{-t s_2Q^2_\bullet}\right)\d s_1.
\endalign $$

\bf 4.1. Local section \rm \newline

Let $\sigma_a$, $a=1,2,3$, be the Pauli matrices. 
Antihermitian generators of su(2) are $k^a:=-\frac{i}{2}\sigma_a$ for which $\left[k^a,k^b\right]=\epsilon_{abc}k^c$.
Denoting ${\bold k}:=(k^1,k^2,k^3)$ and ${\bold n\cdot\bold k}:={n_a k^a}$,
any element of $SU(2)$ can be written in the form $e^{\fii{\bold n\cdot \bold k}}=\cos\frac{\fii}{2}+2\sin\frac{\fii}{2}{\bold n\cdot \bold k}$ where
$\fii\in[0,2\pi]$ and ${\bold n}=(n_1,n_2,n_3)\in S^2$.

For all $\fii\in[0,2\pi)$ and ${\bold n}\in S^2$ define
$$
[0,1]\ni y\mapsto s_{\fii,{\bold n}}(y):=e^{y\fii{\bold n\cdot \bold k}}\in SU(2), \tag4.4
$$
so that $(\fii,{\bold n})\mapsto s_{\fii,{\bold n}}$ gives a local section $s$ of $\Cal A\to SU(2)$
over $U_+ = SU(2)\setminus \{-1\},$
the vector potential associated to the open path $(\fii,{\bold n})\mapsto s_{\fii,{\bold n}}$ is
$A=s^{-1} ds=  \fii{\bold n\cdot \bold k}\, dy.$

\bf 4.2. Results \rm \newline

Let $K$ be a compact (Hausdorff) space, and let $C(K)$ be a Banach space of continuous functions $w:\,K\to\C$ equipped with the sup norm.
By the Riesz representation theorem, the topological dual $M(K)$ of $C(K)$ can be considered as a set of regular complex Borel measures on $K$.
Equip $M(K)$ with the weak-star topology.
In this topology, we say that a net $\{\mu_i\}_{i\in\Cal I}\subseteq M(K)$, where $\Cal I$ is a directed set, converges to a point
$\mu\in M(K)$ if $\lim_{i\in\Cal I}\int w\d\mu_i=\int w\d\mu$ for any $w\in C(K)$; we denote $\mu=\mathop{\hbox{w$^*$-lim}}_{i\in\Cal I}\mu_i$.

Let ${\bold{area}}(S^2)\in M(S^2)$ denote the measure defined by the area 2-form of $S^2$, 
$\delta_a\in M\big([0,2\pi]\big)$ the Dirac measure concentrated on $a\in(0,2\pi)$, and
$$
\fii_j^k:=2\pi\frac{2j+1}{4\tilde k}=2\pi\frac{2j+1}{k+2}\in(0,2\pi).
$$
Recall that $k$ is the level of the irreducible representation 
of the loop algebra $L\gm$ on $H_b$, $\tilde k=(k+2)/4$, and
$2j= 0,1,2,..., k$  labels the irreducible representations of $SU(2)$ on the
vacuum sector.

\proclaim{Theorem A}
When $s^*\Theta^t[1]$ is interpreted as a measure on $[0,2\pi]$,
$$
\mathop{\hbox{\rm w$^*$-lim}}_{t\to\infty}s^*\Theta^t[1]=-\sqrt{\pi}\delta_{\fii_j^k}.
$$
Especially, the limit does not depend on $u$ and $\alpha$.
\endproclaim
\demo{Proof} 
See appendix. 
\enddemo
\noindent
To get an integral form, $\Theta^t[1]$ must be multiplied by $\pi^{-1/2}$. 

From this we conclude that the normalized correction term in (3.13) to the 3-form part in the Chern character becomes
$$ \frac{k+2}{8\pi^2}     (\sin \fii -\fii )  \delta_{\fii_j^k}\otimes{\bold{area}}(S^2). \tag4.6$$
This follows from the fact that the 2-form potential $\theta_+$ on $U_+$ of the twist form
$$\Omega= - (k+2) \frac{1}{4\pi^2} \sin^2 (\fii/2)\, d\fii \wedge {\bold{area}}(S^2) \tag4.7$$ is equal to 
$$ \theta_+ = \frac{k+2}{8\pi^2} (\sin \fii -\fii  ) \, {\bold{area}}(S^2).\tag4.8$$

 \proclaim{Theorem B} 
When $s^*\Theta^t[3]$ is interpreted as a measure on $[0,2\pi]\times S^2$
$$
\mathop{\hbox{\rm w$^*$-lim}}_{t\to\infty}s^*\Theta^t[3]=-\sqrt{\pi}i\left(j+\frac{1}{2}\right)\delta_{\fii_j^k}\otimes{\bold{area}}(S^2).
$$
Especially, the limit does not depend on $u$ and $\alpha$.
\endproclaim
\demo{Proof} 
See appendix. 
\enddemo
\noindent
For an integral form, we multiply $\Theta^t[3]$ by $(\sqrt{\pi}i)^{-1}(2\pi)^{-1}$. 
The result agrees
with the calculations of invariants in the $SU(2)$ case in [Mi2], using different methods.  

 Since the normalized form
has support only on the $S^2$ surface $\fii=\fii_j^k,$ it is actually globally defined integral
form on $SU(2),$  whereas the correction term (4.6) defines a nonintegral global form.

It is interesting to note that, if $\text{supp}\,\alpha\subseteq[1-\epsilon,1]$, $\epsilon>0$, $\epsilon\approx 0$, and $u=1$ then
$$\align
(\sqrt{\pi}i)^{-1}(2\pi)^{-1}\mathop{\hbox{\rm w$^*$-lim}}_{t\to\infty}s^*\Theta_1^t[3]&=\frac{k+2}{8\pi^2}\left[\sin \fii-2\fii+{\Cal O}(\epsilon)\right] \delta_{\fii_j^k}\otimes{\bold{area}}(S^2), \\
(\sqrt{\pi}i)^{-1}(2\pi)^{-1}\mathop{\hbox{\rm w$^*$-lim}}_{t\to\infty}s^*\Theta_2^t[3]&=\frac{k+2}{8\pi^2}\left[\fii-\sin \fii+{\Cal O}(\epsilon)\right] \delta_{\fii_j^k}\otimes{\bold{area}}(S^2),
\endalign $$
so that normalized $s^*\Theta_2^t[3]$ annihilates the correction term (4.6) in the limit $t\to\infty$
(for the proof, see appendix, equations A.1--4).

\vskip 0.3in

\bf Discussion \rm 

\vskip 0.2in
The question remains how much information one can get from the Chern character in the 
case of an arbitrary compact simple Lie group.  One expects that in the limit $t\to \infty$ the 
forms are supported on the conjugacy class in $G$ corresponding to zeros of the supercharge
$Q_A.$ These can be described as follows (and they were evaluated in the second paper of
[FHT]). The elements of $G$ can be parametrized  as $g=\exp(2\pi a),$ where $a$ is in the Lie algebra
of $G.$ The constant vector potential $A=a$ on the circle $S^1$ has the holonomy $g.$ 
Since the construction of the operators $Q_A$  is equivariant with respect to ((constant) gauge 
transformations we may assume that $a$ is in the Cartan subalgebra $\bold h$ of $G.$ 
Actually, performing an additional gauge transformation one can require that 
$$(a+  \overset{\vee}\to{d} , \alpha_i) \leq 0 $$
where $\alpha_i$'s are the simple roots of the affine Lie algebra based on $G,$   $\vee$  means the duality transformation $\bold h\to \bold h^*$ 
defined by the Killing form, and $(\cdot,\cdot)$ is the inner product in $\bold h^*$ coming from the 
Killing form. Here $d=-i \frac{d}{d\theta}$ is the derivation of the loop algebra. 
Then the only zeros of $Q_A$ are in the weight subspace $\lambda + \rho$ where $\lambda$ is
the highest weight of the $G$ representation on the bosonic vacuum sector, $\rho$ is half the sum
of the positive roots (which is also a highest weight on the fermionic vacuum sector). The value of $a$ 
corresponding to the nonzero kernel of $Q_A$ is given by
$$ \tilde k \overset{\vee}\to{a} =  -\lambda -\rho. $$

\vskip 0.3in

\bf References \rm 

\vskip 0.2in
[At] M. Atiyah: K theory past and present.  math.KT/0012213.
Sitzungsberichte der Berliner  Mathematischen Gesellschaft,  411--417, Berliner Math. Gesellschaft, Berlin, 2001.

[AS] M. Atiyah and G. Segal:  Twisted K-theory. math.KT/0407054
 
[Bi] J.-M. Bismut: Localization formulas, superconnections, and the index
 theorem for families.   Comm. Math. Phys.  \bf 103, \rm  no. 1, 127--166 (1986) 

[BCMMS] P. Bouwknegt, A. Carey,  V. Mathai,  M. K. Murray, and D. Stevenson: Twisted K-theory and K-theory of
bundle gerbes.  0106194. Commun. Math. Phys.  \bf 228, \rm  17-49 (2002)

[CMM] A. Carey, J. Mickelsson, and M.K. Murray: Index theory, gerbes,
and hamiltonian quantization hep-th/9511151
Commun.Math.Phys.\bf 183, \rm 707-722 (1997)

[Do] Christopher L. Douglas:  On the twisted K-homology of simple Lie groups.  math.AT/0402082

[FdV]  H. Freudenthal and H. de Vries: \it Linear Lie Groups. \rm Pure Appl. Math. \bf 35, \rm Academic Press, 
New York (1969)

[Fr]  D. Freed: Twisted K-theory and loop groups. math.AT/0206237. Publ. in the proceedings of
ICM2002, Beijing.

[FHT] D. Freed, M. Hopkins, and C. Teleman: Twisted  equivariant K-theory with complex coefficients.
math.AT/0206257. Twisted K-theory and loop group representations.  math.AT/0312155

[Mi1]  J. Mickelsson: Gerbes, (twisted) K theory, and the supersymmetric WZW model. hep-th/0206139.
Publ. in  \it Infinite Dimensional Groups and Manifolds,   \rm  ed. by T. Wurzbacher. IRMA Lectures in
Mathematics and Theoretical Physics \bf 5, \rm Walter de Gruyter \ , Berlin (2004)

[Mi2]  J. Mickelsson: Twisted K theory invariants.  math.AT/0401130. Lett. in Math. Phys. \bf 71, \rm  109-121 (2005)

[Qu] D. Quillen: Superconnections and the Chern character.  Topology
\bf 24, \rm  no. 1, 89--95 (1985)

\enddocument

\vskip 0.3in
\bf Appendix: Zeros of $Q_A$ \rm 

\vskip 0.2in

The zero modes of the operator $Q_A$ are the same as its square $Q_A^2$ but the latter is easier to handle.
First we gauge transform the potential $A$ to a constant potential in the Cartan subalgebra $\hm$ and denote by
$\mu\in \hm^*$ its dual with respect to the Killing form. Then 
$$ Q_A^2 = h + \frac{N}{24} + |\tilde k \mu|^2 + 2\tilde k \mu_i h^{i}   \tag{A1}.$$

Acting with $h$ on the highest weight vectors, corresponding to the weight $\lambda +\rho$  in $H_b\otimes H_f,$
gives the eigenvalue  $|\lambda +\rho|^2 -|\rho|^2.$ Here $\rho$ is half the sum of positive roots of $\gm.$ 
In our normalization of the inner product,
$|\rho|^2 = N/24,$ [FdV], and therefore the eigenvalue of  $Q_A^2$ on the highest weight vector is 
$$ Q_A^2 (\lambda) = |\lambda +\rho|^2 + \tilde k^2|\mu|^2 + 2\tilde k (\mu, \lambda +\rho).\tag{A2}$$

When acting on a general vector of weight $\lambda'$ in the representation space the eigenvalue of $h$ is shifted by
$\tilde k(\lambda' -\lambda - \rho)(2d)$ and so the eigenvalue of $Q_A^2$ becomes
$$ Q_A^2(\lambda') =  |\lambda +\rho|^2 + \tilde k^2 |\mu|^2+ 2\tilde k (\mu, \lambda') + 2\tilde (\lambda' -\lambda -\rho)(d).
\tag{A3}$$
We can rewrite this expression as 
$$Q_A^2 (\lambda') = |\lambda +\rho + \tilde k \mu|^2 -+2\tilde k (\lambda' -\lambda -\rho, \mu + \text{\v{$d$}} ), \tag{A4}$$
where \v{$d$} is the dual root of $d.$ 
After action by a Weyl group element of the affine Lie algebra (implemented here by a gauge transformation) we may assume 
that $\mu + \text{\v{$d$}}$ is in the  Weyl  chamber defined by
$$ (\mu + \text{\v{$d$}}, \alpha_i) \leq 0,$$ 
where $\alpha_i$ with $i=0,1,2,\dots \ell$ are the simple roots. 
Since the difference $\lambda +\rho -\lambda'$ is a sum of simple roots, we conclude that the second term in (A4) is nonnegative.
Thus in order that the eigenvalue $Q_A^2(\lambda')=0$ the first term has to vanish. This implies
$$\tilde k \mu = - \lambda -\rho. \tag{A5}$$

From the last equation follows that all the coordinates $(\mu,\alpha_i)$ are strictly negative , by $(\rho,\alpha_i)=1,$ and so 
the second term in (A4) vanishes if and only if $\lambda' -\lambda -\rho = 0.$ We conclude that the only zeros of $Q_A$ are in
the vacuum $\lambda'=\lambda + \rho$ when $\mu +\text{\v{$d$}}$ is in the negative of the fundamental Weyl chamber  and then 
$\mu$ is given by (A5).

Denote the superconnection $\sqrt{t}\sigma Q_\bullet + \delta + \hat\omega_c  - \frac{\sigma}{4\sqrt{t}}\langle\psi,F\rangle$ by $D_{1,t}$
and define a new superconnection $D_{2,t}:=\sqrt{t}\sigma Q_\bullet + \delta$.
Choose a one parameter family of superconnections joining them: redefine
$$D_t:=u D_{1,t}+(1-u) D_{2,t}=\sqrt{t}\sigma Q_\bullet + \delta + u\hat\omega_c  - \frac{u\sigma}{4\sqrt{t}}\langle\psi,F\rangle. \tag4.1$$
where $u\in[0,1]$ is here considered as a constant, i.e., $\delta u=0$.
Thus,
$$\align
D_t^2=& t Q_\bullet^2 +\sqrt{t}\sigma\underbrace{\Big(-\delta Q_\bullet+u[Q_\bullet,\hat\omega_c]\Big)}_{=:B_1\,\hbox{(1-form)}}+
\underbrace{\Big(u\delta\hat\omega_c+u^2\hat\omega_c^2-\frac{u}{4}\{Q_\bullet,\langle\psi,F\rangle\}\Big)}_{=:B_2\,\hbox{(2-form)}}
\\
&-\frac{\sigma}{4\sqrt{t}}\underbrace{\Big(-u\delta\langle\psi,F\rangle+u^2[\langle\psi,F\rangle,\hat\omega_c]\Big)}_{=:B_3\,\hbox{(3-form)}} 
=t\big(Q_\bullet^2-E_t\big) \tag4.2
\endalign $$
where
$$
E_t:=-\frac{1}{t}\left(\sqrt{t}\sigma B_1+B_2-\frac{\sigma}{4\sqrt{t}} B_3\right). \tag4.3
$$

Using the perturbation series expansion 
$$
\tr^\sigma e^{-t(Q^2_\bullet-E_t)}=\tr^\sigma e^{-t Q^2_\bullet}+\sum_{n=1}^\infty t^n\int_{\Delta_n}\tr^\sigma\left(e^{-t s_1Q^2_\bullet}E_te^{-t s_2Q^2_\bullet}
E_t\cdots e^{-t s_{n+1}Q^2_\bullet}\right)\d s_1\cdots\d s_n,
$$
where $\Delta_n$ is the standard $n$-simplex
$$
\left\{(s_1,s_2,\ldots,s_{n+1})\in\R^{n+1}\,\bigg|\,0\leq s_a\leq 1,\;\sum_{a=1}^{n+1} s_a=1\right\},
$$
one gets
$$\align
\Theta^t=&\tr^\sigma e^{-t Q^2_\bullet}
+t\int_{\Delta_1}\tr^\sigma\left(e^{-t s_1Q^2_\bullet}E_te^{-t s_2Q^2_\bullet}\right)\d s_1
\\
&+t^2\int_{\Delta_2}\tr^\sigma\left(e^{-t s_1Q^2_\bullet}E_te^{-t s_2Q^2_\bullet}E_t e^{-t s_3 Q^2_\bullet}\right)\d s_1\d s_2 
\\
&+t^3\int_{\Delta_3}\tr^\sigma\left(e^{-t s_1Q^2_\bullet}E_te^{-t s_2Q^2_\bullet}E_t e^{-t s_3 Q^2_\bullet}E_te^{-t s_4 Q^2_\bullet}\right)\d s_1\d s_2\d s_3 
\\
=&\Theta^t[1]+\Theta^t[3]
\endalign $$
where $\Theta^t[3]:=\Theta_1^t[3]+\Theta_2^t[3]+\Theta_3^t[3]$ and
$$\align
\Theta^t[1]:=&{-\sqrt{t}\int_{\Delta_1}\tr\left(e^{-t s_1Q^2_\bullet}B_1e^{-t s_2Q^2_\bullet}\right)\d s_1},\\
\Theta_1^t[3]:=&t\sqrt{t}\int_{\Delta_3}\tr\left(e^{-t s_1Q^2_\bullet}B_1 e^{-t s_2Q^2_\bullet}B_1 e^{-t s_3 Q^2_\bullet}B_1 e^{-t s_4 Q^2_\bullet}\right)\d s_1\d s_2\d s_3,\\
\Theta_2^t[3]:=&\sqrt{t}\int_{\Delta_2}\tr\left(e^{-t s_1Q^2_\bullet}B_1e^{-t s_2Q^2_\bullet}B_2 e^{-t s_3 Q^2_\bullet}
+e^{-t s_1Q^2_\bullet}B_2e^{-t s_2Q^2_\bullet}B_1 e^{-t s_3 Q^2_\bullet}\right)\d s_1\d s_2,\\
\Theta_3^t[3]:=&\frac{1}{4\sqrt{t}}\int_{\Delta_1}\tr\left(e^{-t s_1Q^2_\bullet} B_3 e^{-t s_2Q^2_\bullet}\right)\d s_1.
\endalign $$

\bf 4.1. Local section \rm \newline

Let $\sigma_a$, $a=1,2,3$, be the Pauli matrices. 
Antihermitian generators of su(2) are $k^a:=-\frac{i}{2}\sigma_a$ for which $\left[k^a,k^b\right]=\epsilon_{abc}k^c$.
Denoting ${\bold k}:=(k^1,k^2,k^3)$ and ${\bold n\cdot\bold k}:={n_a k^a}$,
any element of $SU(2)$ can be written in the form $e^{\fii{\bold n\cdot \bold k}}=\cos\frac{\fii}{2}+2\sin\frac{\fii}{2}{\bold n\cdot \bold k}$ where
$\fii\in[0,2\pi]$ and ${\bold n}=(n_1,n_2,n_3)\in S^2$.

For all $\fii\in[0,2\pi)$ and ${\bold n}\in S^2$ define
$$
[0,1]\ni y\mapsto s_{\fii,{\bold n}}(y):=e^{y\fii{\bold n\cdot \bold k}}\in SU(2), \tag4.4
$$
so that $(\fii,{\bold n})\mapsto s_{\fii,{\bold n}}$ gives a local section $s$ of $\Cal A\to SU(2)$
over $U_+ = SU(2)\setminus \{-1\},$
the vector potential associated to the open path $(\fii,{\bold n})\mapsto s_{\fii,{\bold n}}$ is
$A=s^{-1} ds=  \fii{\bold n\cdot \bold k}\, dy.$

\bf 4.2. Results \rm \newline

Let $K$ be a compact (Hausdorff) space, and let $C(K)$ be a Banach space of continuous functions $w:\,K\to\C$ equipped with the sup norm.
By the Riesz representation theorem, the topological dual $M(K)$ of $C(K)$ can be considered as a set of regular complex Borel measures on $K$.
Equip $M(K)$ with the weak-star topology.
In this topology, we say that a net $\{\mu_i\}_{i\in\Cal I}\subseteq M(K)$, where $\Cal I$ is a directed set, converges to a point
$\mu\in M(K)$ if $\lim_{i\in\Cal I}\int w\d\mu_i=\int w\d\mu$ for any $w\in C(K)$; we denote $\mu=\mathop{\hbox{w$^*$-lim}}_{i\in\Cal I}\mu_i$.

Let ${\bold{area}}(S^2)\in M(S^2)$ denote the measure defined by the area 2-form of $S^2$, 
$\delta_a\in M\big([0,2\pi]\big)$ the Dirac measure concentrated on $a\in(0,2\pi)$, and
$$
\fii_j^k:=2\pi\frac{2j+1}{4\tilde k}=2\pi\frac{2j+1}{k+2}\in(0,2\pi).
$$
Recall that $k$ is the level of the irreducible representation 
of the loop algebra $L\gm$ on $H_b$, $\tilde k=(k+2)/4$, and
$2j= 0,1,2,..., k$  labels the irreducible representations of $SU(2)$ on the
vacuum sector.

\proclaim{Theorem A}
When $s^*\Theta^t[1]$ is interpreted as a measure on $[0,2\pi]$,
$$
\mathop{\hbox{\rm w$^*$-lim}}_{t\to\infty}s^*\Theta^t[1]=-\sqrt{\pi}\delta_{\fii_j^k}.
$$
Especially, the limit does not depend on $u$ and $\alpha$.
\endproclaim
\demo{Proof} 
See appendix. 
\enddemo
\noindent
To get an integral form, $\Theta^t[1]$ must be multiplied by $\pi^{-1/2}$. 

From this we conclude that the normalized correction term in (3.13) to the 3-form part in the Chern character becomes
$$ \frac{k+2}{8\pi^2}     (\sin \fii -\fii )  \delta_{\fii_j^k}\otimes{\bold{area}}(S^2). \tag4.6$$
This follows from the fact that the 2-form potential $\theta_+$ on $U_+$ of the twist form
$$\Omega= - (k+2) \frac{1}{4\pi^2} \sin^2 (\fii/2)\, d\fii \wedge {\bold{area}}(S^2) \tag4.7$$ is equal to 
$$ \theta_+ = \frac{k+2}{8\pi^2} (\sin \fii -\fii  ) \, {\bold{area}}(S^2).\tag4.8$$

 \proclaim{Theorem B} 
When $s^*\Theta^t[3]$ is interpreted as a measure on $[0,2\pi]\times S^2$
$$
\mathop{\hbox{\rm w$^*$-lim}}_{t\to\infty}s^*\Theta^t[3]=-\sqrt{\pi}i\left(j+\frac{1}{2}\right)\delta_{\fii_j^k}\otimes{\bold{area}}(S^2).
$$
Especially, the limit does not depend on $u$ and $\alpha$.
\endproclaim
\demo{Proof} 
See appendix. 
\enddemo
\noindent
For an integral form, we multiply $\Theta^t[3]$ by $(\sqrt{\pi}i)^{-1}(2\pi)^{-1}$. 
The result agrees
with the calculations of invariants in the $SU(2)$ case in [Mi2], using different methods.  

 Since the normalized form
has support only on the $S^2$ surface $\fii=\fii_j^k,$ it is actually globally defined integral
form on $SU(2),$  whereas the correction term (4.6) defines a nonintegral global form.

It is interesting to note that, if $\text{supp}\,\alpha\subseteq[1-\epsilon,1]$, $\epsilon>0$, $\epsilon\approx 0$, and $u=1$ then
$$\align
(\sqrt{\pi}i)^{-1}(2\pi)^{-1}\mathop{\hbox{\rm w$^*$-lim}}_{t\to\infty}s^*\Theta_1^t[3]&=\frac{k+2}{8\pi^2}\left[\sin \fii-2\fii+{\Cal O}(\epsilon)\right] \delta_{\fii_j^k}\otimes{\bold{area}}(S^2), \\
(\sqrt{\pi}i)^{-1}(2\pi)^{-1}\mathop{\hbox{\rm w$^*$-lim}}_{t\to\infty}s^*\Theta_2^t[3]&=\frac{k+2}{8\pi^2}\left[\fii-\sin \fii+{\Cal O}(\epsilon)\right] \delta_{\fii_j^k}\otimes{\bold{area}}(S^2),
\endalign $$
so that normalized $s^*\Theta_2^t[3]$ annihilates the correction term (4.6) in the limit $t\to\infty$
(for the proof, see appendix, equations A.1--4).

\vskip 0.3in

\bf Discussion \rm 

\vskip 0.2in
The question remains how much information one can get from the Chern character in the 
case of an arbitrary compact simple Lie group.  One expects that in the limit $t\to \infty$ the 
forms are supported on the conjugacy class in $G$ corresponding to zeros of the supercharge
$Q_A.$ These can be described as follows (and they were evaluated in the second paper of
[FHT]). The elements of $G$ can be parametrized  as $g=\exp(2\pi a),$ where $a$ is in the Lie algebra
of $G.$ The constant vector potential $A=a$ on the circle $S^1$ has the holonomy $g.$ 
Since the construction of the operators $Q_A$  is equivariant with respect to ((constant) gauge 
transformations we may assume that $a$ is in the Cartan subalgebra $\bold h$ of $G.$ 
Actually, performing an additional gauge transformation one can require that 
$$(a+  \overset{\vee}\to{d} , \alpha_i) \leq 0 $$
where $\alpha_i$'s are the simple roots of the affine Lie algebra based on $G,$   $\vee$  means the duality transformation $\bold h\to \bold h^*$ 
defined by the Killing form, and $(\cdot,\cdot)$ is the inner product in $\bold h^*$ coming from the 
Killing form. Here $d=-i \frac{d}{d\theta}$ is the derivation of the loop algebra. 
Then the only zeros of $Q_A$ are in the weight subspace $\lambda + \rho$ where $\lambda$ is
the highest weight of the $G$ representation on the bosonic vacuum sector, $\rho$ is half the sum
of the positive roots (which is also a highest weight on the fermionic vacuum sector). The value of $a$ 
corresponding to the nonzero kernel of $Q_A$ is given by
$$ \tilde k \overset{\vee}\to{a} =  -\lambda -\rho. $$

 
\vskip 0.3in

\bf Appendix: The proofs of theorems A and B \rm 

\vskip 0.2in
{\bf Pullbacks.} First we calculate some pullbacks of forms with respect to the local section $s$.
By a direct calculation one gets, over $U_+,$
$$\align
s^*\omega_{\fii,{\bold n}}(y)=&\left[y-\alpha(y)\right]\,k^an_a\,\d\fii+\left\{\sin(y\fii)-\alpha(y)\left[\sin(y\fii)+\sin\big((1-y)\fii\big)\right]\right\}\,k^a\d n_a\\
&+\left\{\cos(y\fii)-1-\alpha(y)\left[\cos(y\fii)-\cos\big((1-y)\fii\big)\right]\right\}\epsilon_{abc}k^a n_b \d n_c
\endalign $$

Let $\{e_m\}_{m\in\Z}$, $e_m(y):=e^{2\pi i m y}$, be the Fourier basis of $L^2\big(S^1,\d y\big)$.
Then for all $\fii\in(0,2\pi)$ the following series converge pointwise at $y\in (0,1)$:
$$\align
\alpha(y)&=\sum_{m\in\Z}\alpha_m e_m(y)\hbox{ where $\{\alpha_m\}_{m}$ is rapidly decreasing and $\alpha_{-m}=\overline{\alpha_m}$},
\\
y&=\frac{1}{2}+\frac{i}{2\pi}\sum_{m\in\Z\atop m\ne0}\frac{1}{m}e_m(y),
\\
\sin(y\fii)&=\sum_{m\in\Z}\frac{\fii(\cos\fii-1)+2\pi i m\sin\fii}{(2\pi m)^2-\fii^2}e_m(y),
\\
\cos(y\fii)&=\sum_{m\in\Z}\frac{-\fii\sin\fii+2\pi i m(\cos\fii-1)}{(2\pi m)^2-\fii^2}e_m(y).
\endalign$$

In quantization of $\omega$ one replaces $k^a e_m$ by $\sqrt{2}S_m^a$. The pull-back of $\hat\omega$ is then
$$\align
s^*\hat\omega_{\fii,{\bold n}}&=\sqrt{2}\left\{
\left(\frac{1}{2}-\alpha_0\right)n_aS_0^a+
\sum_{m\ne0}\left(\frac{i}{2\pi m}-\alpha_m\right)n_aS_m^a\right\}\d\fii 
\\
&+\sqrt{2}\sum_{m\in\Z}\left[\frac{\fii(\cos\fii-1)+2\pi i m\sin\fii}{(2\pi m)^2-\fii^2}+2\fii(1-\cos\fii)\sum_{l\in\Z}\frac{\alpha_{m-l}}{(2\pi l)^2-\fii^2}\right]S_m^a\d n_a 
\\
&-\sqrt{2}\epsilon_{abc}S_0^a n_b\d n_c 
\\
+\sqrt{2}\sum_{m\in\Z}&\left[\frac{-\fii\sin\fii+2\pi i m(\cos\fii-1)}{(2\pi m)^2-\fii^2}+4\pi i(1-\cos\fii)\sum_{l\in\Z}\frac{l\alpha_{m-l}}{(2\pi l)^2-\fii^2}\right]
\epsilon_{abc}S_m^a n_b\d n_c.
\endalign $$
Also
$$\align
s^*Q_\bullet|_{\fii,\bold n}&=Q+\sqrt{2}\tilde k\Fii n_a\psi_0^a, \tag4.5a
\\
(s^*Q_\bullet|_{\fii,\bold n})^2&=Q^2+2\sqrt{2}i\tilde k \Fii n_a S_0^a+2{\tilde k}^2\Fii^2, \tag4.5b
\\
s^*\delta Q_\bullet|_{\fii,\bold n}&=\sqrt{2}\tilde k n_a\psi_0^a\d\Fii+\sqrt{2}\tilde k\Fii \psi_0^a\d n_a. \tag4.5c
\endalign $$
where $\Fii=\fii/(2\pi)$ is the normalized angle.

{\bf The term $\Theta_1^t[3]$.} By direct calculation, one gets
$$
s^*(\delta Q_\bullet-u[Q_\bullet,\hat\omega_c])|_{\fii,\bold n}=\sqrt{2}\tilde k C_{m,a}\psi_m^a
$$
where for all $m\in\Z$ 
$$\align
C_{m,1}(\fii,{\bold n}):=&n_1d_m(\fii)\d\fii+\left[e_m(\fii)+\left(n_2^2+n_3^2\right)f_m(\fii)\right]\d n_1
\\
&+\left[n_3g_m(\fii)-n_1n_2 f_m(\fii)\right]\d n_2+\left[-n_2g_m(\fii)-n_1n_3 f_m(\fii)\right]\d n_3,
\\
C_{m,2}(\fii,{\bold n}):=&n_2d_m(\fii)\d\fii+\left[-n_3g_m(\fii)-n_1n_2 f_m(\fii)\right]\d n_1
\\
&+\left[e_m(\fii)+\left(n_1^2+n_3^2\right)f_m(\fii)\right]\d n_2+\left[n_1g_m(\fii)-n_2n_3 f_m(\fii)\right]\d n_3,
\\
C_{m,3}(\fii,{\bold n}):=&n_3d_m(\fii)\d\fii+\left[n_2g_m(\fii)-n_1n_3 f_m(\fii)\right]\d n_1
\\
&+\left[-n_1g_m(\fii)-n_2n_3 f_m(\fii)\right]\d n_2+\left[e_m(\fii)+\left(n_1^2+n_2^2\right)f_m(\fii)\right]\d n_3,
\endalign $$
and 
$$\align
a_m(\fii)&:=\frac{\fii(\cos\fii-1)+2\pi i m\sin\fii}{(2\pi m)^2-\fii^2}+2\fii(1-\cos\fii)j_m(\fii),
\\
b_m(\fii)&:=\frac{-\fii\sin\fii+2\pi i m(\cos\fii-1)}{(2\pi m)^2-\fii^2}+4\pi(1-\cos\fii)k_m(\fii),
\\
d_m(\fii)&:=-\frac{u}{2\pi} - i u m \alpha_m,\;\;m\ne0,\;\;\;\;
d_0(\fii):=\frac{1}{2\pi},
\\
e_m(\fii)&:=i u m a_m(\fii),\;\;m\ne0,\;\;\;\;
e_0(\fii):=\frac{\fii}{2\pi},
\\
f_m(\fii)&:=\frac{u \fii}{2\pi}b_m(\fii),\;\;m\ne0,\;\;\;\;
f_0(\fii):=\frac{u \fii}{2\pi}\big(b_0(\fii)-1\big),
\\
g_m(\fii)&:=\frac{u \fii}{2\pi}a_m(\fii)-i u m b_m(\fii),
\\
{j_m(\fii)}&:=\sum_{l\in\Z}\frac{\alpha_{m-l}}{(2\pi l)^2-\fii^2},
\\
{k_m(\fii)}&:=\sum_{l\in\Z}\frac{il\alpha_{m-l}}{(2\pi l)^2-\fii^2}
\endalign $$
for all $m\in\Z$ and $\fii\in(0,2\pi)$. The terms $C_{m,j}$ are smooth at $\fii=0.$ 

Denoting $s_n^a:=\sqrt{2}S_n^a$ and ${\bold s}:=(s_0^1,s_0^2,s_0^3)$ and using equation
$[h,\psi_n^a]=2\tilde kn\psi_n^a$ one gets
$$\align
&s^*\Theta_1^t[3]_{\fii,\bold n}-\tilde k^3(2t)^{3/2}C_{n,a}(\fii,{\bold n})\wedge C_{m,b}(\fii,{\bold n})\wedge C_{l,c}(\fii,{\bold n})\times
\\
&\times\int_{\Delta_{3}}\tr\Big(e^{-t s_1(h+2i\tilde k \Fii {\bold n\cdot\bold s}+2{\tilde k}^2\Fii^2)}\psi_n^a
e^{-t s_2(h+2i\tilde k \Fii {\bold n\cdot\bold s}+2{\tilde k}^2\Fii^2)}\psi_m^b \times
\\
&\times e^{-t s_3(h+2i\tilde k \Fii {\bold n\cdot\bold s}+2{\tilde k}^2\Fii^2)} 
\psi_l^c e^{-t s_{4}(h+2i\tilde k \Fii {\bold n\cdot\bold s}+2{\tilde k}^2\Fii^2)}\Big)\d^3s
\\
&=-\tilde k^3(2t)^{3/2}C_{n,a}(\fii,{\bold n})\wedge C_{m,b}(\fii,{\bold n})\wedge C_{l,c}(\fii,{\bold n})
\int_{\Delta_{3}}e^{-2\tilde k t[s_1(n+m+l)+s_2(m+l)+s_3 l]}\times
\\
&\times
\tr\Big(e^{-t s_12i\tilde k \Fii {\bold n\cdot\bold s}}\psi_n^a
e^{-t s_2 2i\tilde k \Fii {\bold n\cdot s}}\psi_m^b e^{-t s_3 2i\tilde k \Fii {\bold n\cdot \bold s}} 
\psi_l^c e^{-t s_4 2i\tilde k \Fii {\bold n\cdot\bold s}} 
e^{-t(h+2{\tilde k}^2\Fii^2)}\Big)\d^3s
\endalign $$
Next we divide the above equation into three parts.

{\bf Trace part.}
Denoting traces over $H_f$ and $H_b$ by $\tr_f$ and $\tr_b$, respectively, we get
$$\align
&\tr\Big(e^{-t s_12i\tilde k \Fii {\bold n\cdot\bold s}}\psi_n^a
e^{-t s_2 2i\tilde k \Fii {\bold n\cdot\bold s}}\psi_m^b e^{-t s_3 2i\tilde k \Fii {\bold n\cdot\bold s}} 
\psi_l^c e^{-t s_4 2i\tilde k \Fii {\bold n\cdot\bold s}} e^{-t h}\Big)
\\
&=e^{-t/8}\tr_b\Big(e^{-t 2i\tilde k \Fii {\bold n\cdot\bold t}}e^{-t h_b}\Big)\times
\\
&\times\tr_f\Big(e^{-t s_12i\tilde k \Fii {\bold n\cdot\underline{\bold k}}}\psi_n^a
e^{-t s_2 2i\tilde k \Fii {\bold n\cdot\underline{\bold k}}}\psi_m^b e^{-t s_3 2i\tilde k \Fii {\bold n\cdot\underline{\bold k}}} 
\psi_l^c e^{-t s_4 2i\tilde k \Fii {\bold n\cdot\underline{\bold k}}} e^{-t2\tilde k h_f}\Big)
\endalign $$
where $t_n^a:=\sqrt{2}T_n^a$, ${\bold t}:=(t_0^1,t_0^2,t_0^3)$, $k_n^a:=\sqrt{2}K_n^a$, and $\underline{\bold k}:=(k_0^1,k_0^2,k_0^3)$.

Since $[t^a_0,T^b_n]=\epsilon_{abc}T^c_n$ and $[k^a_0,\psi^b_n,]=\epsilon_{abc}\psi^c_n$ 
one gets for any $\theta\in\R$,
$$
e^{-i\theta\bold n\cdot\bold t }T_n^a e^{i\theta\bold n\cdot\bold t}=R({\bold n},\theta)_{a,b}T_n^b,\;\;\;\;
e^{-i\theta\bold n\cdot\underline{\bold k} }\psi_n^a e^{i\theta\bold n\cdot\underline{\bold k}}=R({\bold n},\theta)_{a,b}\psi_n^b
$$
where
$$
\align
& R({\bold n},\theta):= \\
& \pmatrix
n_1^2+(1-n_1^2)\cosh\theta & n_1n_2(1-\cosh\theta)-in_3\sinh\theta & n_3n_1(1-\cosh\theta)+in_2\sinh\theta \\
n_1n_2(1-\cosh\theta)+in_3\sinh\theta & n_2^2+(1-n_2^2)\cosh\theta & n_2n_3(1-\cosh\theta)-in_1\sinh\theta \\
n_3n_1(1-\cosh\theta)-in_2\sinh\theta & n_2n_3(1-\cosh\theta)+in_1\sinh\theta & n_3^2+(1-n_3^2)\cosh\theta
\endpmatrix.
\endalign
$$
Let $\tr_{f,n}$ and $\tr_{b,n}$ denote traces calculated over the eigenspace of operators $2\tilde k h_f$ and $h_b-j(j+1)/2$ associated to
an eigenvalue $2\tilde k n$, $n\in\{0,1,2,...\}$, respectively.
Then 
$$
\tr_b\Big(e^{-t 2i\tilde k \Fii {\bold n\cdot\bold t}}e^{-t h_b}\Big)=e^{-t j(j+1)/2}\sum_{n=0}^\infty e^{-t 2\tilde k n}\tr_{b,n}\Big(e^{-t 2i\tilde k \Fii {\bold n\cdot\bold t}}\Big).
$$ 
Since $e^{-i\theta\bold n\cdot\bold t }T_n^a=R({\bold n},\theta)_{a,b}T_n^b e^{-i\theta\bold n\cdot\bold t}$ and the biggest eigenvalue of $e^{-i\theta\bold n\cdot\bold t}$ operating
on the bosonic vacuum is $e^{\theta j}$, in the limit $t\to\infty$, $e^{-t 2\tilde k n}\tr_{b,n}\Big(e^{-t 2i\tilde k \Fii {\bold n\cdot\bold t}}\Big)$ is (at most) of order
$e^{jt2\tilde k\Fii}e^{nt2\tilde k(\Fii-1)}$. Hence, $\tr_b\Big(e^{-t 2i\tilde k \Fii {\bold n\cdot\bold t}}e^{-t h_b}\Big)$ is asymptotic to
$e^{t(2\tilde k\Fii j-j(j+1)/2)}$.

Similarly,
$$\align
&\tr_f\Big(e^{-t s_12i\tilde k \Fii {\bold n\cdot\underline{\bold k}}}\psi_n^a
e^{-t s_2 2i\tilde k \Fii {\bold n\cdot\underline{\bold k}}}\psi_m^b e^{-t s_3 2i\tilde k \Fii {\bold n\cdot\underline{\bold k}}} 
\psi_l^c e^{-t s_4 2i\tilde k \Fii {\bold n\cdot\underline{\bold k}}} e^{-t2\tilde k h_f}\Big) 
\\
&=\sum_{p=0}^\infty e^{-t 2\tilde k p}\tr_{f,p}\Big(e^{-t s_12i\tilde k \Fii {\bold n\cdot\underline{\bold k}}}\psi_n^a
e^{-t s_2 2i\tilde k \Fii {\bold n\cdot\underline{\bold k}}}\psi_m^b e^{-t s_3 2i\tilde k \Fii {\bold n\cdot\underline{\bold k}}} 
\psi_l^c e^{-t s_4 2i\tilde k \Fii {\bold n\cdot\underline{\bold k}}}\Big).
\endalign $$
Immediately, one sees that if $n+m+l\ne 0$, the above trace vanishes. 
By direct calculation,
$$\align
&\tr_{f,p}\Big(e^{-t s_12i\tilde k \Fii {\bold n\cdot\underline{\bold k}}}\psi_n^a
e^{-t s_2 2i\tilde k \Fii {\bold n\cdot\underline{\bold k}}}\psi_m^b e^{-t s_3 2i\tilde k \Fii {\bold n\cdot\underline{\bold k}}} 
\psi_l^c e^{-t s_4 2i\tilde k \Fii {\bold n\cdot\underline{\bold k}}}\Big)
\\
&=\tr_{f,p}\Big(e^{-t s_12i\tilde k \Fii {\bold n\cdot\underline{\bold k}}}e^{-t s_2 2i\tilde k \Fii {\bold n\cdot\underline{\bold k}}} R({\bold n},-t s_2 2\tilde k \Fii)_{a,\tilde a}
\psi_n^{\tilde a}
\psi_m^b R({\bold n},t s_3 2\tilde k \Fii)_{c,\tilde c}\times
\\
&\times\psi_l^{\tilde c} 
e^{-t s_3 2i\tilde k \Fii {\bold n\cdot\underline{\bold k}}} e^{-t s_4 2i\tilde k \Fii {\bold n\cdot\underline{\bold k}}}\Big)
\\
&=R({\bold n},-t s_2 2\tilde k \Fii)_{a,\tilde a} R({\bold n},t s_3 2\tilde k \Fii)_{c,\tilde c}
\tr_{f,p}\Big(\psi_n^{\tilde a}\psi_m^b\psi_l^{\tilde c} 
e^{-t 2i\tilde k \Fii {\bold n\cdot\underline{\bold k}}}\Big).
\endalign $$
As before, in the limit $t\to\infty$, $e^{-t 2\tilde k p}\tr_{f,p}\Big(\psi_n^{\tilde a}\psi_m^b\psi_l^{\tilde c} 
e^{-t 2i\tilde k \Fii {\bold n\cdot\underline{\bold k}}}\Big)$ is (at most) of order
$e^{t\tilde k\Fii}e^{pt2\tilde k(\Fii-1)}$. 
Thus we get
$$\align
&\tr_f\Big(e^{-t s_12i\tilde k \Fii {\bold n\cdot\underline{\bold k}}}\psi_n^a
e^{-t s_2 2i\tilde k \Fii {\bold n\cdot\underline{\bold k}}}\psi_m^b e^{-t s_3 2i\tilde k \Fii {\bold n\cdot\underline{\bold k}}} 
\psi_l^c e^{-t s_4 2i\tilde k \Fii {\bold n\cdot\underline{\bold k}}} e^{-t2\tilde k h_f}\Big) 
\\
&\sim R({\bold n},-t s_2 2\tilde k \Fii)_{a,\tilde a} R({\bold n},t s_3 2\tilde k \Fii)_{c,\tilde c}
\tr_{f,0}\Big(\psi_n^{\tilde a}\psi_m^b\psi_l^{\tilde c} e^{-t 2i\tilde k \Fii {\bold n\cdot\underline{\bold k}}}\Big).
\endalign$$
To get a non-vanishing trace, indexes must satisfy one of the following conditions:
$$\align
n=m=l=0,  \tilde a\ne b\ne \tilde c\ne\tilde a, \;\; &\tr_{f,0}\Big(\psi_0^{\tilde a}\psi_0^b\psi_0^{\tilde c} e^{-t 2i\tilde k \Fii {\bold n\cdot\underline{\bold k}}}\Big)
= 2i\epsilon_{\tilde a b\tilde c}\cosh(t\tilde k\Fii), 
\\
n=m=l=0,  \tilde a=b, \;\; &\tr_{f,0}\Big(\psi_0^b\psi_0^b\psi_0^{\tilde c} e^{-t 2i\tilde k \Fii {\bold n\cdot\underline{\bold k}}}\Big)
=-2n_{\tilde c}\sinh(t\tilde k\Fii), 
\\
n=m=l=0,  b=\tilde c,\;\; &\tr_{f,0}\Big(\psi_0^{\tilde a}\psi_0^b\psi_0^b e^{-t 2i\tilde k \Fii {\bold n\cdot\underline{\bold k}}}\Big)
 = -2n_{\tilde a}\sinh(t\tilde k\Fii), 
\\
n=m=l=0,  \tilde a=\tilde c\ne b, \;\; &\tr_{f,0}\Big(\psi_0^{\tilde a}\psi_0^b\psi_0^{\tilde a} e^{-t 2i\tilde k \Fii {\bold n\cdot\underline{\bold k}}}\Big)
= 2n_b\sinh(t\tilde k\Fii), 
\\
n=-s,\,m=s,\,l=0,  \tilde a=b, \;\; &\tr_{f,0}\Big(\psi_{-s}^b\psi_s^b\psi_0^{\tilde c} e^{-t 2i\tilde k \Fii {\bold n\cdot\underline{\bold k}}}\Big)
=-4n_{\tilde c}\sinh(t\tilde k\Fii), 
\\
n=0,\,m=-s,\,l=s,  b=\tilde c,\;\; &\tr_{f,0}\Big(\psi_0^{\tilde a}\psi_{-s}^b\psi_s^b e^{-t 2i\tilde k \Fii {\bold n\cdot\underline{\bold k}}}\Big)
 = -4n_{\tilde a}\sinh(t\tilde k\Fii), 
\\
n=-s,\,m=0,\,l=s,  \tilde a=\tilde c, \;\; &\tr_{f,0}\Big(\psi_{-s}^{\tilde a}\psi_0^b\psi_s^{\tilde a} e^{-t 2i\tilde k \Fii {\bold n\cdot\underline{\bold k}}}\Big)
= 4n_b\sinh(t\tilde k\Fii), 
\endalign $$
where $s>0$, and we have used equation
$$
\tr_{f,0}\Big(\psi_0^a 
e^{-t 2i\tilde k \Fii {\bold n\cdot\underline{\bold k}}}\Big)\tr_{\C^2}\Big(\sigma_a 
e^{-t 2i\tilde k \Fii {\bold n\cdot\bold k}}\Big)=-2n_a\sinh(t\tilde k\Fii).
$$

{\bf Integral part.} In this subsection, we calculate 
$$\align
A^{\Fii,{\bold n};t}_{n,m,l}(a,b,c):=&
\sum_{\tilde a,\tilde c=1}^3\tr_{f,0}\Big(\psi_n^{\tilde a}\psi_m^b\psi_l^{\tilde c}e^{-t 2i\tilde k \Fii {\bold n\cdot\underline{\bold k}}}\Big)\times\\
&\times\int_{\Delta_{3}}e^{-2\tilde k t[s_1(n+m+l)+s_2(m+l)+s_3 l]}R({\bold n},-t s_2 2\tilde k \Fii)_{a,\tilde a} 
R({\bold n},t s_3 2\tilde k \Fii)_{c,\tilde c}\d^3s.
\endalign $$
Taking account of index conditions, the integral can be reduced to the sum of the following integrals:
$$\align
I(\alpha,\beta)&:=\int_{\Delta_{3}}e^{\alpha s_2+\beta s_3}\d^3s=\int_0^1\int_0^{1-s_1}\int_0^{1-s_1-s_2}e^{\alpha s_2+\beta s_3}\d s_3\d s_2\d s_1\\
&=\frac{\beta^2(e^\alpha-\alpha-1)-\alpha^2(e^\beta-\beta-1)}{\alpha^2(\alpha-\beta)\beta^2}
\endalign $$
where $\alpha:=2\tilde kt(\Fii u -m-l)$, $\beta:=2\tilde kt(\Fii v-l)$, and $u$, $v\in\{-1,0,1\}$.
After long simplification, we get
$$\align
A^{\Fii,{\bold n};t}_{0,0,0}(a,b,c)&=-\frac{1}{6}n_1n_2n_3e^{t \tilde k \Fii}
+\frac{e^{t \tilde k \Fii}}{4\Fii\tilde k t} \left[2 n_1 n_2 n_3 + i\epsilon_{abc} (1-n_b^2)\right]+{\Cal O}\left(t^{-2}e^{t \tilde k \Fii}\right),\\
&\text{when}\;a\ne b\ne c\ne a,\\
A^{\Fii,{\bold n};t}_{-s,s,0}(a,b,c)&= 
\frac{e^{t \tilde k \Fii}}{2\tilde k t}
\left[n_an_bn_c\frac{\Fii^2}{s(s^2-\Fii^2)}-i\epsilon_{abd}n_dn_c\frac{\Fii}{s^2-\Fii^2}-\delta_{ab}n_c\frac{s}{s^2-\Fii^2}\right]\\
&+{\Cal O}\left(t^{-2}e^{t \tilde k \Fii}\right),\\
A^{\Fii,{\bold n};t}_{0,-s,s}(a,b,c)&= 
\frac{e^{t \tilde k \Fii}}{2\tilde k t}
\left[n_an_bn_c\frac{\Fii^2}{s(s^2-\Fii^2)}-i\epsilon_{bcd}n_dn_a\frac{\Fii}{s^2-\Fii^2}-\delta_{bc}n_a\frac{s}{s^2-\Fii^2}\right]\\
&+{\Cal O}\left(t^{-2}e^{t \tilde k \Fii}\right),\\
A^{\Fii,{\bold n};t}_{-s,0,s}(a,b,c)&={\Cal O}\left(t^{-2}e^{t \tilde k \Fii}\right),
\endalign $$
when $t\to\infty$.

{\bf Form part}.
Next we calculate form
$C_{0,a} (\fii,{\bold n})\wedge C_{0,b} (\fii,{\bold n})\wedge C_{0,c} (\fii,{\bold n})$.
First, we note that 
$$
C_{0,a} (\fii,{\bold n})\wedge C_{0,b} (\fii,{\bold n})\wedge C_{0,c} (\fii,{\bold n})
=\epsilon_{abc}C_{0,1} (\fii,{\bold n})\wedge C_{0,2} (\fii,{\bold n})\wedge C_{0,3} (\fii,{\bold n}).
$$
Using condition $\d n_a\wedge\d n_b\wedge\d n_c=0$ one gets
$$
C_{0,1} (\fii,{\bold n})\wedge C_{0,2} (\fii,{\bold n})\wedge C_{0,3} (\fii,{\bold n})
=d_0 (\fii)\left[\big(e_0 (\fii)+f_0 (\fii)\big)^2+\big(g_0 (\fii)\big)^2\right]\d\fii\wedge{\bold{area}}(S^2)|_{\bold n}
$$
where ${\bold{area}}(S^2)|_{\bold n}:=n_1\d n_2\wedge\d n_3+n_2\d n_3\wedge\d n_1+n_3\d n_1\wedge\d n_2$ is the area 2-form of $S^2$ at $\bold n$.
In spherical coordinates $(\theta,\phi)$, for which
$n_1=\sin\theta\cos\phi$, $n_2=\sin\theta\sin\phi$, and $n_3=\cos\theta$, 2-form ${\bold{area}}(S^2)|_{\theta,\phi}=\sin\theta\,\d\theta\wedge\d\phi$.

After collecting the results of the previous subsections together, we arrive at 
$$
s^*\Theta_1^t[3]_{\fii,\bold n}-\tilde k^2(2t)^{1/2}e^{-t[2{\tilde k}^2\Fii^2-(2j+1)\tilde k\Fii +j(j+1)/2+1/8]}\left[\Omega _1(\fii,{\bold n})+\Omega _2(\fii,{\bold n})+{\Cal O}(t^{-1})\right]
$$
where
$$\align
\Omega _1(\fii,{\bold n})&:=\frac{2\tilde k t}{e^{t \tilde k \Fii}}A^{\Fii,{\bold n};t}_{0,0,0}(a,b,c)
C_{0,a} (\fii,{\bold n})\wedge C_{0,b} (\fii,{\bold n})\wedge C_{0,c} (\fii,{\bold n}), \\
\Omega _2(\fii,{\bold n})&:=\frac{2\tilde k t}{e^{t \tilde k \Fii}}\sum_{s=1}^\infty\big[
A^{\Fii,{\bold n};t}_{-s,s,0}(a,b,c)C_{-s,a} (\fii,{\bold n})\wedge C_{s,b} (\fii,{\bold n})\wedge C_{0,c} (\fii,{\bold n})\\
&+A^{\Fii,{\bold n};t}_{0,-s,s}(a,b,c)C_{0,a} (\fii,{\bold n})\wedge C_{-s,b} (\fii,{\bold n})\wedge C_{s,c} (\fii,{\bold n})\big].
\endalign $$
After long but straightforward calculations, one gets
$$\align
\Omega_1(\fii,{\bold n})&=\frac{2 i}{\Fii}d_0 (\fii)\left[\big(e_0 (\fii)+f_0 (\fii)\big)^2+\big(g_0 (\fii)\big)^2\right]\d\fii\wedge{\bold{area}}(S^2)|_{\bold n}\\
&=\frac{i}{\pi}\Fii\left[(1-u)^2+u^2\big(a_0 (\fii)^2+b_0 (\fii)^2\big)+2(1-u)ub_0 (\fii)\right]\d\fii\wedge{\bold{area}}(S^2)|_{\bold n},\\
\Omega _2(\fii,{\bold n})&=4 i d_0 (\fii)\sum_{s=1}^\infty
\frac{\Fii\left[|e_s (\fii)+f_s (\fii)|^2+|g_s (\fii)|^2\right]+
2\,\text{Im}\Big(s\big(\overline{e_s (\fii)}+\overline{f_s (\fii)}\big)g_s (\fii)\Big)}{(\Fii+s)(\Fii-s)}\times\\
&\times
\d\fii\wedge{\bold{area}}(S^2)|_{\bold n}\\
&=\frac{2i}{\pi}u^2\sum_{s=1}^\infty\left[\Fii\left(|a_s (\fii)|^2+|b_s (\fii)|^2\right)
-2 s\,\text{Im}\left(a_s (\fii)b_{-s} (\fii)\right)\right]\d\fii\wedge{\bold{area}}(S^2)|_{\bold n},
\endalign $$
which shows that 
$$
\Omega _1(\fii,{\bold n})+\Omega _2(\fii,{\bold n})=\frac{i}{\pi}F_u(\fii)\,\d\fii\wedge{\bold{area}}(S^2)|_{\bold n}
$$
where
$$\align
F_u(\fii):=&\frac{\fii}{2\pi}\left[(1-u)^2+2(1-u)ub_0 (\fii)\right]\\
&+u^2\sum_{s\in\Z}\left[\frac{\fii}{2\pi}\left(|a_s (\fii)|^2+|b_s (\fii)|^2\right)
-2 s\,\text{Im}\left(a_s (\fii)b_{-s} (\fii)\right)\right].
\endalign $$
To conclude,
$$
s^*\Theta_1^t[3]_{\fii,\bold n}-\tilde k^2(2t)^{1/2}e^{-t[2{\tilde k}^2\Fii^2-(2j+1)\tilde k\Fii +j(j+1)/2+1/8]}\left[
\frac{i}{\pi}F_u(\fii)\,\d\fii\wedge{\bold{area}}(S^2)|_{\bold n}+{\Cal O}(t^{-1})\right] 
\tag{A.1}
$$
Note that if we assume that $\text{supp}\,\alpha\subseteq[1-\epsilon,1]$ where $\epsilon>0$ and $\epsilon\approx 0$, then
$$
F_u(\fii)=\frac{\fii}{2\pi}\left[(1-u)^2+2(1-u)u\frac{\sin\fii}{\fii}+u^2\left(2-\frac{\sin\fii}{\fii}\right)\right]+{\Cal O}(\epsilon).
\tag{A.2}
$$

{\bf The term $\Theta_2^t[3]$.}
Denote $s^*B_2|_{\fii,\bold n}=T_{n,b}S_n^b+T \cdot {\bold 1}$ where $T_{n,b}$ and $T$ are 2-forms
and ${\bold 1}$ is the identity operator.
Then
$$
s^*\Theta_2^t[3]_{\fii,\bold n}=-\sqrt{2t}\tilde k C_{m,a}\wedge T_{n,b}E_{m,n}(a,b)-\sqrt{2t}\tilde k C_{m,a}\wedge T\,
\tr\Big(\psi_m^a e^{-t(h+2i\tilde k \Fii {\bold n\cdot\bold s}+2{\tilde k}^2\Fii^2)}\Big)
$$
where
$$\align
&E_{m,n}(a,b):=\int_{\Delta_{2}}\tr\Big(e^{-t s_1(h+2i\tilde k \Fii {\bold n\cdot\bold s}+2{\tilde k}^2\Fii^2)}\psi_m^a
e^{-t s_2(h+2i\tilde k \Fii {\bold n\cdot\bold s}+2{\tilde k}^2\Fii^2)}S_n^b e^{-t s_3(h+2i\tilde k \Fii {\bold n\cdot\bold s}+2{\tilde k}^2\Fii^2)}\\
&+e^{-t s_1(h+2i\tilde k \Fii {\bold n\cdot\bold s}+2{\tilde k}^2\Fii^2)}S_n^b
e^{-t s_2(h+2i\tilde k \Fii {\bold n\cdot\bold s}+2{\tilde k}^2\Fii^2)}\psi_m^ae^{-t s_3(h+2i\tilde k \Fii {\bold n\cdot\bold s}+2{\tilde k}^2\Fii^2)}\Big)\d s_1\d s_2\\
&=\delta_{n,-m}e^{-t2\tilde k^2\Fii^2}\int_{\Delta_{2}}\Big[
R({\bold n},-t s_2 2\tilde k \Fii)_{a,\tilde a}e^{2\tilde k t s_2 m}
\tr\Big(\psi_m^{\tilde a}S_{-m}^be^{-t 2i\tilde k \Fii {\bold n\cdot\bold s}}e^{-t h}\big)\\
&+R({\bold n},t s_2 2\tilde k \Fii)_{a,\tilde a}e^{-2\tilde k t s_2 m}
\tr\Big(S_{-m}^b\psi_m^{\tilde a}e^{-t 2i\tilde k \Fii {\bold n\cdot\bold s}}e^{-t h}\big)\Big]\d s_1\d s_2.
\endalign $$
Since $S_{-m}^b=T_{-m}^b+K_{-m}^b$, $S_{-m}^b\psi_m^{\tilde a}=\psi_m^{\tilde a}S_{-m}^b-\lambda_{\tilde a b c}\psi_0^c$,
$\tr_b\Big(T_{-m}^be^{-t 2i\tilde k \Fii {\bold n\cdot\bold t}}e^{-t h_b}\Big)\sim\delta_{m,0}2^{-1/2}i j n_b e^{t2\tilde k\Fii j}e^{-t j(j+1)/2}$
we get
$$
E_{m,n}(a,b)\sim-\delta_{m,0}\delta_{n,0}e^{-t[2{\tilde k}^2\Fii^2-(2j+1)\tilde k\Fii +j(j+1)/2+1/8]}\frac{i}{\sqrt{2}}\left(j+\frac{1}{2}\right)n_a n_b.
$$
In addition,
$$\align
&C _{m,a}\tr\left(\psi_m^ae^{-t[h_b+2\tilde k h_f+\frac{1}{8}+2 i\tilde k \Fii({\bold{n\cdot t+n\cdot\underline{k}}})
+2{\tilde k}^2\Fii^2]}\right)\\
&\sim 
-e^{-t[2{\tilde k}^2\Fii^2-(2j+1)\tilde k\Fii+j(j+1)/2+1/8]}\underbrace{n_aC _{0,a}}_{=\d\Fii},
\endalign $$
so that
$$
s^*\Theta_2^t[3]_{\fii,\bold n}\sim
\frac{i}{\pi}\tilde k^2\sqrt{2t}e^{-t[2{\tilde k}^2\Fii^2-(2j+1)\tilde k\Fii+j(j+1)/2+1/8]}\d\fii\wedge\left(\frac{2j+1}{4\tilde k}\frac{n_b T_{0,b}}{\sqrt{2}}+\frac{T}{2 i\tilde k}\right).
\tag{A.3}
$$
Collecting significant terms
$$\align
&s^*\delta\hat\omega_c|_{\fii,\bold n}=\sqrt{2}\big(b_0(\fii)-1\big)S_0^a\epsilon_{abc}\d n_b\wedge\d n_c+\ldots,\\
&s^*\hat\omega^2_c|_{\fii,\bold n}=\frac{1}{\sqrt{2}}S_0^d\epsilon_{d a\tilde a}\big[
\big(a_s(\fii)\d n_a+b_s(\fii)\epsilon_{abc}n_b\d n_c\big)\wedge
\big(a_{-s}(\fii)\d n_{\tilde a}+b_{-s}(\fii)\epsilon_{\tilde a\tilde b\tilde c}n_{\tilde b}\d n_{\tilde c}\big)\\
&+\epsilon_{abc}\epsilon_{\tilde a\tilde b\tilde c}n_bn_{\tilde b}\d n_c\wedge \d n_{\tilde c}
-2\big(a_0(\fii)\d n_a+b_0(\fii)\epsilon_{abc}n_b\d n_c\big)\wedge\epsilon_{\tilde a\tilde b\tilde c}n_{\tilde b}\d n_{\tilde c}\big]\\
&+\tilde k s\big(a_s(\fii)\d n_a+b_s(\fii)\epsilon_{abc}n_b\d n_c\big)\wedge
\big(a_{-s}(\fii)\d n_{a}+b_{-s}(\fii)\epsilon_{a\tilde b\tilde c}n_{\tilde b}\d n_{\tilde c}\big)+\ldots,\\
&s^*\{Q_\bullet,\langle\psi,F\rangle\}_{\fii,\bold n}=2 i \tilde F_{-n}^b S_n^b+2\sqrt{2}\tilde k\Fii n_b\tilde F_0^bI,
\endalign $$
where $\tilde F_{-n}^b$ is a Fourier coefficient of $s^*\langle\psi,F\rangle_{\fii,\bold n}$,
we see that
$$\align
\frac{n_b T_{0,b}}{\sqrt{2}}&=\left\{u^2\left[\sum_{s\in\Z}\left(|a_s(\fii)|^2+|b_s(\fii)|^2\right)+1-2b_0(\fii)\right]
+2u\big(b_0(\fii)-1\big)\right\}{\bold{area}}(S^2)|_{\bold n}\\
&-\frac{u}{2\sqrt{2}}in_b\tilde F_0^b+\ldots,\\
T&=u^2\tilde k\sum_{s\in\Z}(-2 s)\big(a_s(\fii)b_{-s}(\fii)-a_{-s}(\fii)b_s(\fii)\big){\bold{area}}(S^2)|_{\bold n}
-\frac{u}{2}\sqrt{2}\tilde k\Fii n_b\tilde F_0^b+\ldots.
\tag{A.4}
\endalign 
$$

{\bf Conclusions.}
Using the formal notations $\delta(\fii-a)$ for the Dirac measure $\delta_a$
concentrated on $a\in(0,2\pi)$ and $w(\fii)$ for the measure defined by a 1-form $w(\fii)\d\fii$ where $g\in C\big([0,2\pi]\big)$, one gets
$$
\delta(\fii-a)=\mathop{\hbox{w$^*$-lim}}_{r\to\infty}\sqrt{\frac{r}{\pi}}e^{-r(\fii-a)^2}
$$
and for any $p\ge 1$,
$$
\mathop{\hbox{w$^*$-lim}}_{r\to\infty}\frac{1}{r^p}\sqrt{r}\,e^{-r\fii^2}=0.
$$
Since $2{\tilde k}^2\Fii^2-(2j+1)\tilde k\Fii +j(j+1)/2+1/8=\left(\frac{\tilde k}{\sqrt{2}\pi}\right)^2(\fii-\fii_j^k)^2$ where 
$$
\fii_j^k:=2\pi\frac{2j+1}{4\tilde k}=2\pi\frac{2j+1}{k+2}\in(0,2\pi)
$$
and using equatoins (A.1), (A.3), and (A.4),
we may conclude that,
when $s^*\Theta_1^t[3]$ and $s^*\Theta_2^t[3]$ are interpreted as measures on $[0,2\pi]\times S^2$,
$$
\mathop{\hbox{\rm w$^*$-lim}}_{t\to\infty}\left(s^*\Theta_1^t[3]+s^*\Theta_2^t[3]\right)=-\sqrt{\pi}i\left(j+\frac{1}{2}\right)\delta_{\fii_j^k}\otimes{\bold{area}}(S^2).
$$
Proceeding similarly as above, one sees immediately that the weak-star limit of the three-form $s^*\Theta^t_3[3]$
can be interpreted as the zero measure. Thus,
$$
\mathop{\hbox{\rm w$^*$-lim}}_{t\to\infty}s^*\Theta^t[3]=-\sqrt{\pi}i\left(j+\frac{1}{2}\right)\delta_{\fii_j^k}\otimes{\bold{area}}(S^2)
$$
and theorem B follows.

For large $t,$
$$
s^*\Theta^t[1]_{\fii,\bold n}\sim -\sqrt{2t}\tilde k 
e^{-t[2{\tilde k}^2\Fii^2-(2j+1)\tilde k\Fii+j(j+1)/2+1/8]}{\d\Fii}.
$$
Hence, when $s^*\Theta^t[1]$ is interpreted as a measure on $[0,2\pi]$,
$$
\mathop{\hbox{\rm w$^*$-lim}}_{t\to\infty}s^*\Theta^t[1]=-\sqrt{\pi}\delta_{\fii_j^k}.
$$
This proves theorem A.
 

\vskip 0.3in

\bf References \rm 

\vskip 0.2in
[At] M. Atiyah: K theory past and present.  math.KT/0012213.
Sitzungsberichte der Berliner  Mathematischen Gesellschaft,  411--417, Berliner Math. Gesellschaft, Berlin, 2001.

[AS] M. Atiyah and G. Segal:  Twisted K-theory. math.KT/0407054
 
[Bi] J.-M. Bismut: Localization formulas, superconnections, and the index
 theorem for families.   Comm. Math. Phys.  \bf 103, \rm  no. 1, 127--166 (1986) 

[BCMMS] P. Bouwknegt, A. Carey,  V. Mathai,  M. K. Murray, and D. Stevenson: Twisted K-theory and K-theory of
bundle gerbes.  0106194. Commun. Math. Phys.  \bf 228, \rm  17-49 (2002)

[CMM] A. Carey, J. Mickelsson, and M.K. Murray: Index theory, gerbes,
and hamiltonian quantization hep-th/9511151
Commun.Math.Phys.\bf 183, \rm 707-722 (1997)

[Do] Christopher L. Douglas:  On the twisted K-homology of simple Lie groups.  math.AT/0402082

[FdV]  H. Freudenthal and H. de Vries: \it Linear Lie Groups. \rm Pure Appl. Math. \bf 35, \rm Academic Press, 
New York (1969)

[Fr]  D. Freed: Twisted K-theory and loop groups. math.AT/0206237. Publ. in the proceedings of
ICM2002, Beijing.

[FHT] D. Freed, M. Hopkins, and C. Teleman: Twisted  equivariant K-theory with complex coefficients.
math.AT/0206257. Twisted K-theory and loop group representations.  math.AT/0312155

[Mi1]  J. Mickelsson: Gerbes, (twisted) K theory, and the supersymmetric WZW model. hep-th/0206139.
Publ. in  \it Infinite Dimensional Groups and Manifolds,   \rm  ed. by T. Wurzbacher. IRMA Lectures in
Mathematics and Theoretical Physics \bf 5, \rm Walter de Gruyter \ , Berlin (2004)

[Mi2]  J. Mickelsson: Twisted K theory invariants.  math.AT/0401130. Lett. in Math. Phys. \bf 71, \rm  109-121 (2005)

[Qu] D. Quillen: Superconnections and the Chern character.  Topology
\bf 24, \rm  no. 1, 89--95 (1985)

\enddocument

 
\vskip 0.3in

\bf Appendix: The proofs of theorems A and B \rm 

\vskip 0.2in
{\bf Pullbacks.} First we calculate some pullbacks of forms with respect to the local section $s$.
By a direct calculation one gets, over $U_+,$
$$\align
s^*\omega_{\fii,{\bold n}}(y)=&\left[y-\alpha(y)\right]\,k^an_a\,\d\fii+\left\{\sin(y\fii)-\alpha(y)\left[\sin(y\fii)+\sin\big((1-y)\fii\big)\right]\right\}\,k^a\d n_a\\
&+\left\{\cos(y\fii)-1-\alpha(y)\left[\cos(y\fii)-\cos\big((1-y)\fii\big)\right]\right\}\epsilon_{abc}k^a n_b \d n_c
\endalign $$

Let $\{e_m\}_{m\in\Z}$, $e_m(y):=e^{2\pi i m y}$, be the Fourier basis of $L^2\big(S^1,\d y\big)$.
Then for all $\fii\in(0,2\pi)$ the following series converge pointwise at $y\in (0,1)$:
$$\align
\alpha(y)&=\sum_{m\in\Z}\alpha_m e_m(y)\hbox{ where $\{\alpha_m\}_{m}$ is rapidly decreasing and $\alpha_{-m}=\overline{\alpha_m}$},
\\
y&=\frac{1}{2}+\frac{i}{2\pi}\sum_{m\in\Z\atop m\ne0}\frac{1}{m}e_m(y),
\\
\sin(y\fii)&=\sum_{m\in\Z}\frac{\fii(\cos\fii-1)+2\pi i m\sin\fii}{(2\pi m)^2-\fii^2}e_m(y),
\\
\cos(y\fii)&=\sum_{m\in\Z}\frac{-\fii\sin\fii+2\pi i m(\cos\fii-1)}{(2\pi m)^2-\fii^2}e_m(y).
\endalign$$

In quantization of $\omega$ one replaces $k^a e_m$ by $\sqrt{2}S_m^a$. The pull-back of $\hat\omega$ is then
$$\align
s^*\hat\omega_{\fii,{\bold n}}&=\sqrt{2}\left\{
\left(\frac{1}{2}-\alpha_0\right)n_aS_0^a+
\sum_{m\ne0}\left(\frac{i}{2\pi m}-\alpha_m\right)n_aS_m^a\right\}\d\fii 
\\
&+\sqrt{2}\sum_{m\in\Z}\left[\frac{\fii(\cos\fii-1)+2\pi i m\sin\fii}{(2\pi m)^2-\fii^2}+2\fii(1-\cos\fii)\sum_{l\in\Z}\frac{\alpha_{m-l}}{(2\pi l)^2-\fii^2}\right]S_m^a\d n_a 
\\
&-\sqrt{2}\epsilon_{abc}S_0^a n_b\d n_c 
\\
+\sqrt{2}\sum_{m\in\Z}&\left[\frac{-\fii\sin\fii+2\pi i m(\cos\fii-1)}{(2\pi m)^2-\fii^2}+4\pi i(1-\cos\fii)\sum_{l\in\Z}\frac{l\alpha_{m-l}}{(2\pi l)^2-\fii^2}\right]
\epsilon_{abc}S_m^a n_b\d n_c.
\endalign $$
Also
$$\align
s^*Q_\bullet|_{\fii,\bold n}&=Q+\sqrt{2}\tilde k\Fii n_a\psi_0^a, \tag4.5a
\\
(s^*Q_\bullet|_{\fii,\bold n})^2&=Q^2+2\sqrt{2}i\tilde k \Fii n_a S_0^a+2{\tilde k}^2\Fii^2, \tag4.5b
\\
s^*\delta Q_\bullet|_{\fii,\bold n}&=\sqrt{2}\tilde k n_a\psi_0^a\d\Fii+\sqrt{2}\tilde k\Fii \psi_0^a\d n_a. \tag4.5c
\endalign $$
where $\Fii=\fii/(2\pi)$ is the normalized angle.

{\bf The term $\Theta_1^t[3]$.} By direct calculation, one gets
$$
s^*(\delta Q_\bullet-u[Q_\bullet,\hat\omega_c])|_{\fii,\bold n}=\sqrt{2}\tilde k C_{m,a}\psi_m^a
$$
where for all $m\in\Z$ 
$$\align
C_{m,1}(\fii,{\bold n}):=&n_1d_m(\fii)\d\fii+\left[e_m(\fii)+\left(n_2^2+n_3^2\right)f_m(\fii)\right]\d n_1
\\
&+\left[n_3g_m(\fii)-n_1n_2 f_m(\fii)\right]\d n_2+\left[-n_2g_m(\fii)-n_1n_3 f_m(\fii)\right]\d n_3,
\\
C_{m,2}(\fii,{\bold n}):=&n_2d_m(\fii)\d\fii+\left[-n_3g_m(\fii)-n_1n_2 f_m(\fii)\right]\d n_1
\\
&+\left[e_m(\fii)+\left(n_1^2+n_3^2\right)f_m(\fii)\right]\d n_2+\left[n_1g_m(\fii)-n_2n_3 f_m(\fii)\right]\d n_3,
\\
C_{m,3}(\fii,{\bold n}):=&n_3d_m(\fii)\d\fii+\left[n_2g_m(\fii)-n_1n_3 f_m(\fii)\right]\d n_1
\\
&+\left[-n_1g_m(\fii)-n_2n_3 f_m(\fii)\right]\d n_2+\left[e_m(\fii)+\left(n_1^2+n_2^2\right)f_m(\fii)\right]\d n_3,
\endalign $$
and 
$$\align
a_m(\fii)&:=\frac{\fii(\cos\fii-1)+2\pi i m\sin\fii}{(2\pi m)^2-\fii^2}+2\fii(1-\cos\fii)j_m(\fii),
\\
b_m(\fii)&:=\frac{-\fii\sin\fii+2\pi i m(\cos\fii-1)}{(2\pi m)^2-\fii^2}+4\pi(1-\cos\fii)k_m(\fii),
\\
d_m(\fii)&:=-\frac{u}{2\pi} - i u m \alpha_m,\;\;m\ne0,\;\;\;\;
d_0(\fii):=\frac{1}{2\pi},
\\
e_m(\fii)&:=i u m a_m(\fii),\;\;m\ne0,\;\;\;\;
e_0(\fii):=\frac{\fii}{2\pi},
\\
f_m(\fii)&:=\frac{u \fii}{2\pi}b_m(\fii),\;\;m\ne0,\;\;\;\;
f_0(\fii):=\frac{u \fii}{2\pi}\big(b_0(\fii)-1\big),
\\
g_m(\fii)&:=\frac{u \fii}{2\pi}a_m(\fii)-i u m b_m(\fii),
\\
{j_m(\fii)}&:=\sum_{l\in\Z}\frac{\alpha_{m-l}}{(2\pi l)^2-\fii^2},
\\
{k_m(\fii)}&:=\sum_{l\in\Z}\frac{il\alpha_{m-l}}{(2\pi l)^2-\fii^2}
\endalign $$
for all $m\in\Z$ and $\fii\in(0,2\pi)$. The terms $C_{m,j}$ are smooth at $\fii=0.$ 

Denoting $s_n^a:=\sqrt{2}S_n^a$ and ${\bold s}:=(s_0^1,s_0^2,s_0^3)$ and using equation
$[h,\psi_n^a]=2\tilde kn\psi_n^a$ one gets
$$\align
&s^*\Theta_1^t[3]_{\fii,\bold n}-\tilde k^3(2t)^{3/2}C_{n,a}(\fii,{\bold n})\wedge C_{m,b}(\fii,{\bold n})\wedge C_{l,c}(\fii,{\bold n})\times
\\
&\times\int_{\Delta_{3}}\tr\Big(e^{-t s_1(h+2i\tilde k \Fii {\bold n\cdot\bold s}+2{\tilde k}^2\Fii^2)}\psi_n^a
e^{-t s_2(h+2i\tilde k \Fii {\bold n\cdot\bold s}+2{\tilde k}^2\Fii^2)}\psi_m^b \times
\\
&\times e^{-t s_3(h+2i\tilde k \Fii {\bold n\cdot\bold s}+2{\tilde k}^2\Fii^2)} 
\psi_l^c e^{-t s_{4}(h+2i\tilde k \Fii {\bold n\cdot\bold s}+2{\tilde k}^2\Fii^2)}\Big)\d^3s
\\
&=-\tilde k^3(2t)^{3/2}C_{n,a}(\fii,{\bold n})\wedge C_{m,b}(\fii,{\bold n})\wedge C_{l,c}(\fii,{\bold n})
\int_{\Delta_{3}}e^{-2\tilde k t[s_1(n+m+l)+s_2(m+l)+s_3 l]}\times
\\
&\times
\tr\Big(e^{-t s_12i\tilde k \Fii {\bold n\cdot\bold s}}\psi_n^a
e^{-t s_2 2i\tilde k \Fii {\bold n\cdot s}}\psi_m^b e^{-t s_3 2i\tilde k \Fii {\bold n\cdot \bold s}} 
\psi_l^c e^{-t s_4 2i\tilde k \Fii {\bold n\cdot\bold s}} 
e^{-t(h+2{\tilde k}^2\Fii^2)}\Big)\d^3s
\endalign $$
Next we divide the above equation into three parts.

{\bf Trace part.}
Denoting traces over $H_f$ and $H_b$ by $\tr_f$ and $\tr_b$, respectively, we get
$$\align
&\tr\Big(e^{-t s_12i\tilde k \Fii {\bold n\cdot\bold s}}\psi_n^a
e^{-t s_2 2i\tilde k \Fii {\bold n\cdot\bold s}}\psi_m^b e^{-t s_3 2i\tilde k \Fii {\bold n\cdot\bold s}} 
\psi_l^c e^{-t s_4 2i\tilde k \Fii {\bold n\cdot\bold s}} e^{-t h}\Big)
\\
&=e^{-t/8}\tr_b\Big(e^{-t 2i\tilde k \Fii {\bold n\cdot\bold t}}e^{-t h_b}\Big)\times
\\
&\times\tr_f\Big(e^{-t s_12i\tilde k \Fii {\bold n\cdot\underline{\bold k}}}\psi_n^a
e^{-t s_2 2i\tilde k \Fii {\bold n\cdot\underline{\bold k}}}\psi_m^b e^{-t s_3 2i\tilde k \Fii {\bold n\cdot\underline{\bold k}}} 
\psi_l^c e^{-t s_4 2i\tilde k \Fii {\bold n\cdot\underline{\bold k}}} e^{-t2\tilde k h_f}\Big)
\endalign $$
where $t_n^a:=\sqrt{2}T_n^a$, ${\bold t}:=(t_0^1,t_0^2,t_0^3)$, $k_n^a:=\sqrt{2}K_n^a$, and $\underline{\bold k}:=(k_0^1,k_0^2,k_0^3)$.

Since $[t^a_0,T^b_n]=\epsilon_{abc}T^c_n$ and $[k^a_0,\psi^b_n,]=\epsilon_{abc}\psi^c_n$ 
one gets for any $\theta\in\R$,
$$
e^{-i\theta\bold n\cdot\bold t }T_n^a e^{i\theta\bold n\cdot\bold t}=R({\bold n},\theta)_{a,b}T_n^b,\;\;\;\;
e^{-i\theta\bold n\cdot\underline{\bold k} }\psi_n^a e^{i\theta\bold n\cdot\underline{\bold k}}=R({\bold n},\theta)_{a,b}\psi_n^b
$$
where
$$
\align
& R({\bold n},\theta):= \\
& \pmatrix
n_1^2+(1-n_1^2)\cosh\theta & n_1n_2(1-\cosh\theta)-in_3\sinh\theta & n_3n_1(1-\cosh\theta)+in_2\sinh\theta \\
n_1n_2(1-\cosh\theta)+in_3\sinh\theta & n_2^2+(1-n_2^2)\cosh\theta & n_2n_3(1-\cosh\theta)-in_1\sinh\theta \\
n_3n_1(1-\cosh\theta)-in_2\sinh\theta & n_2n_3(1-\cosh\theta)+in_1\sinh\theta & n_3^2+(1-n_3^2)\cosh\theta
\endpmatrix.
\endalign
$$
Let $\tr_{f,n}$ and $\tr_{b,n}$ denote traces calculated over the eigenspace of operators $2\tilde k h_f$ and $h_b-j(j+1)/2$ associated to
an eigenvalue $2\tilde k n$, $n\in\{0,1,2,...\}$, respectively.
Then 
$$
\tr_b\Big(e^{-t 2i\tilde k \Fii {\bold n\cdot\bold t}}e^{-t h_b}\Big)=e^{-t j(j+1)/2}\sum_{n=0}^\infty e^{-t 2\tilde k n}\tr_{b,n}\Big(e^{-t 2i\tilde k \Fii {\bold n\cdot\bold t}}\Big).
$$ 
Since $e^{-i\theta\bold n\cdot\bold t }T_n^a=R({\bold n},\theta)_{a,b}T_n^b e^{-i\theta\bold n\cdot\bold t}$ and the biggest eigenvalue of $e^{-i\theta\bold n\cdot\bold t}$ operating
on the bosonic vacuum is $e^{\theta j}$, in the limit $t\to\infty$, $e^{-t 2\tilde k n}\tr_{b,n}\Big(e^{-t 2i\tilde k \Fii {\bold n\cdot\bold t}}\Big)$ is (at most) of order
$e^{jt2\tilde k\Fii}e^{nt2\tilde k(\Fii-1)}$. Hence, $\tr_b\Big(e^{-t 2i\tilde k \Fii {\bold n\cdot\bold t}}e^{-t h_b}\Big)$ is asymptotic to
$e^{t(2\tilde k\Fii j-j(j+1)/2)}$.

Similarly,
$$\align
&\tr_f\Big(e^{-t s_12i\tilde k \Fii {\bold n\cdot\underline{\bold k}}}\psi_n^a
e^{-t s_2 2i\tilde k \Fii {\bold n\cdot\underline{\bold k}}}\psi_m^b e^{-t s_3 2i\tilde k \Fii {\bold n\cdot\underline{\bold k}}} 
\psi_l^c e^{-t s_4 2i\tilde k \Fii {\bold n\cdot\underline{\bold k}}} e^{-t2\tilde k h_f}\Big) 
\\
&=\sum_{p=0}^\infty e^{-t 2\tilde k p}\tr_{f,p}\Big(e^{-t s_12i\tilde k \Fii {\bold n\cdot\underline{\bold k}}}\psi_n^a
e^{-t s_2 2i\tilde k \Fii {\bold n\cdot\underline{\bold k}}}\psi_m^b e^{-t s_3 2i\tilde k \Fii {\bold n\cdot\underline{\bold k}}} 
\psi_l^c e^{-t s_4 2i\tilde k \Fii {\bold n\cdot\underline{\bold k}}}\Big).
\endalign $$
Immediately, one sees that if $n+m+l\ne 0$, the above trace vanishes. 
By direct calculation,
$$\align
&\tr_{f,p}\Big(e^{-t s_12i\tilde k \Fii {\bold n\cdot\underline{\bold k}}}\psi_n^a
e^{-t s_2 2i\tilde k \Fii {\bold n\cdot\underline{\bold k}}}\psi_m^b e^{-t s_3 2i\tilde k \Fii {\bold n\cdot\underline{\bold k}}} 
\psi_l^c e^{-t s_4 2i\tilde k \Fii {\bold n\cdot\underline{\bold k}}}\Big)
\\
&=\tr_{f,p}\Big(e^{-t s_12i\tilde k \Fii {\bold n\cdot\underline{\bold k}}}e^{-t s_2 2i\tilde k \Fii {\bold n\cdot\underline{\bold k}}} R({\bold n},-t s_2 2\tilde k \Fii)_{a,\tilde a}
\psi_n^{\tilde a}
\psi_m^b R({\bold n},t s_3 2\tilde k \Fii)_{c,\tilde c}\times
\\
&\times\psi_l^{\tilde c} 
e^{-t s_3 2i\tilde k \Fii {\bold n\cdot\underline{\bold k}}} e^{-t s_4 2i\tilde k \Fii {\bold n\cdot\underline{\bold k}}}\Big)
\\
&=R({\bold n},-t s_2 2\tilde k \Fii)_{a,\tilde a} R({\bold n},t s_3 2\tilde k \Fii)_{c,\tilde c}
\tr_{f,p}\Big(\psi_n^{\tilde a}\psi_m^b\psi_l^{\tilde c} 
e^{-t 2i\tilde k \Fii {\bold n\cdot\underline{\bold k}}}\Big).
\endalign $$
As before, in the limit $t\to\infty$, $e^{-t 2\tilde k p}\tr_{f,p}\Big(\psi_n^{\tilde a}\psi_m^b\psi_l^{\tilde c} 
e^{-t 2i\tilde k \Fii {\bold n\cdot\underline{\bold k}}}\Big)$ is (at most) of order
$e^{t\tilde k\Fii}e^{pt2\tilde k(\Fii-1)}$. 
Thus we get
$$\align
&\tr_f\Big(e^{-t s_12i\tilde k \Fii {\bold n\cdot\underline{\bold k}}}\psi_n^a
e^{-t s_2 2i\tilde k \Fii {\bold n\cdot\underline{\bold k}}}\psi_m^b e^{-t s_3 2i\tilde k \Fii {\bold n\cdot\underline{\bold k}}} 
\psi_l^c e^{-t s_4 2i\tilde k \Fii {\bold n\cdot\underline{\bold k}}} e^{-t2\tilde k h_f}\Big) 
\\
&\sim R({\bold n},-t s_2 2\tilde k \Fii)_{a,\tilde a} R({\bold n},t s_3 2\tilde k \Fii)_{c,\tilde c}
\tr_{f,0}\Big(\psi_n^{\tilde a}\psi_m^b\psi_l^{\tilde c} e^{-t 2i\tilde k \Fii {\bold n\cdot\underline{\bold k}}}\Big).
\endalign$$
To get a non-vanishing trace, indexes must satisfy one of the following conditions:
$$\align
n=m=l=0,  \tilde a\ne b\ne \tilde c\ne\tilde a, \;\; &\tr_{f,0}\Big(\psi_0^{\tilde a}\psi_0^b\psi_0^{\tilde c} e^{-t 2i\tilde k \Fii {\bold n\cdot\underline{\bold k}}}\Big)
= 2i\epsilon_{\tilde a b\tilde c}\cosh(t\tilde k\Fii), 
\\
n=m=l=0,  \tilde a=b, \;\; &\tr_{f,0}\Big(\psi_0^b\psi_0^b\psi_0^{\tilde c} e^{-t 2i\tilde k \Fii {\bold n\cdot\underline{\bold k}}}\Big)
=-2n_{\tilde c}\sinh(t\tilde k\Fii), 
\\
n=m=l=0,  b=\tilde c,\;\; &\tr_{f,0}\Big(\psi_0^{\tilde a}\psi_0^b\psi_0^b e^{-t 2i\tilde k \Fii {\bold n\cdot\underline{\bold k}}}\Big)
 = -2n_{\tilde a}\sinh(t\tilde k\Fii), 
\\
n=m=l=0,  \tilde a=\tilde c\ne b, \;\; &\tr_{f,0}\Big(\psi_0^{\tilde a}\psi_0^b\psi_0^{\tilde a} e^{-t 2i\tilde k \Fii {\bold n\cdot\underline{\bold k}}}\Big)
= 2n_b\sinh(t\tilde k\Fii), 
\\
n=-s,\,m=s,\,l=0,  \tilde a=b, \;\; &\tr_{f,0}\Big(\psi_{-s}^b\psi_s^b\psi_0^{\tilde c} e^{-t 2i\tilde k \Fii {\bold n\cdot\underline{\bold k}}}\Big)
=-4n_{\tilde c}\sinh(t\tilde k\Fii), 
\\
n=0,\,m=-s,\,l=s,  b=\tilde c,\;\; &\tr_{f,0}\Big(\psi_0^{\tilde a}\psi_{-s}^b\psi_s^b e^{-t 2i\tilde k \Fii {\bold n\cdot\underline{\bold k}}}\Big)
 = -4n_{\tilde a}\sinh(t\tilde k\Fii), 
\\
n=-s,\,m=0,\,l=s,  \tilde a=\tilde c, \;\; &\tr_{f,0}\Big(\psi_{-s}^{\tilde a}\psi_0^b\psi_s^{\tilde a} e^{-t 2i\tilde k \Fii {\bold n\cdot\underline{\bold k}}}\Big)
= 4n_b\sinh(t\tilde k\Fii), 
\endalign $$
where $s>0$, and we have used equation
$$
\tr_{f,0}\Big(\psi_0^a 
e^{-t 2i\tilde k \Fii {\bold n\cdot\underline{\bold k}}}\Big)\tr_{\C^2}\Big(\sigma_a 
e^{-t 2i\tilde k \Fii {\bold n\cdot\bold k}}\Big)=-2n_a\sinh(t\tilde k\Fii).
$$

{\bf Integral part.} In this subsection, we calculate 
$$\align
A^{\Fii,{\bold n};t}_{n,m,l}(a,b,c):=&
\sum_{\tilde a,\tilde c=1}^3\tr_{f,0}\Big(\psi_n^{\tilde a}\psi_m^b\psi_l^{\tilde c}e^{-t 2i\tilde k \Fii {\bold n\cdot\underline{\bold k}}}\Big)\times\\
&\times\int_{\Delta_{3}}e^{-2\tilde k t[s_1(n+m+l)+s_2(m+l)+s_3 l]}R({\bold n},-t s_2 2\tilde k \Fii)_{a,\tilde a} 
R({\bold n},t s_3 2\tilde k \Fii)_{c,\tilde c}\d^3s.
\endalign $$
Taking account of index conditions, the integral can be reduced to the sum of the following integrals:
$$\align
I(\alpha,\beta)&:=\int_{\Delta_{3}}e^{\alpha s_2+\beta s_3}\d^3s=\int_0^1\int_0^{1-s_1}\int_0^{1-s_1-s_2}e^{\alpha s_2+\beta s_3}\d s_3\d s_2\d s_1\\
&=\frac{\beta^2(e^\alpha-\alpha-1)-\alpha^2(e^\beta-\beta-1)}{\alpha^2(\alpha-\beta)\beta^2}
\endalign $$
where $\alpha:=2\tilde kt(\Fii u -m-l)$, $\beta:=2\tilde kt(\Fii v-l)$, and $u$, $v\in\{-1,0,1\}$.
After long simplification, we get
$$\align
A^{\Fii,{\bold n};t}_{0,0,0}(a,b,c)&=-\frac{1}{6}n_1n_2n_3e^{t \tilde k \Fii}
+\frac{e^{t \tilde k \Fii}}{4\Fii\tilde k t} \left[2 n_1 n_2 n_3 + i\epsilon_{abc} (1-n_b^2)\right]+{\Cal O}\left(t^{-2}e^{t \tilde k \Fii}\right),\\
&\text{when}\;a\ne b\ne c\ne a,\\
A^{\Fii,{\bold n};t}_{-s,s,0}(a,b,c)&= 
\frac{e^{t \tilde k \Fii}}{2\tilde k t}
\left[n_an_bn_c\frac{\Fii^2}{s(s^2-\Fii^2)}-i\epsilon_{abd}n_dn_c\frac{\Fii}{s^2-\Fii^2}-\delta_{ab}n_c\frac{s}{s^2-\Fii^2}\right]\\
&+{\Cal O}\left(t^{-2}e^{t \tilde k \Fii}\right),\\
A^{\Fii,{\bold n};t}_{0,-s,s}(a,b,c)&= 
\frac{e^{t \tilde k \Fii}}{2\tilde k t}
\left[n_an_bn_c\frac{\Fii^2}{s(s^2-\Fii^2)}-i\epsilon_{bcd}n_dn_a\frac{\Fii}{s^2-\Fii^2}-\delta_{bc}n_a\frac{s}{s^2-\Fii^2}\right]\\
&+{\Cal O}\left(t^{-2}e^{t \tilde k \Fii}\right),\\
A^{\Fii,{\bold n};t}_{-s,0,s}(a,b,c)&={\Cal O}\left(t^{-2}e^{t \tilde k \Fii}\right),
\endalign $$
when $t\to\infty$.

{\bf Form part}.
Next we calculate form
$C_{0,a} (\fii,{\bold n})\wedge C_{0,b} (\fii,{\bold n})\wedge C_{0,c} (\fii,{\bold n})$.
First, we note that 
$$
C_{0,a} (\fii,{\bold n})\wedge C_{0,b} (\fii,{\bold n})\wedge C_{0,c} (\fii,{\bold n})
=\epsilon_{abc}C_{0,1} (\fii,{\bold n})\wedge C_{0,2} (\fii,{\bold n})\wedge C_{0,3} (\fii,{\bold n}).
$$
Using condition $\d n_a\wedge\d n_b\wedge\d n_c=0$ one gets
$$
C_{0,1} (\fii,{\bold n})\wedge C_{0,2} (\fii,{\bold n})\wedge C_{0,3} (\fii,{\bold n})
=d_0 (\fii)\left[\big(e_0 (\fii)+f_0 (\fii)\big)^2+\big(g_0 (\fii)\big)^2\right]\d\fii\wedge{\bold{area}}(S^2)|_{\bold n}
$$
where ${\bold{area}}(S^2)|_{\bold n}:=n_1\d n_2\wedge\d n_3+n_2\d n_3\wedge\d n_1+n_3\d n_1\wedge\d n_2$ is the area 2-form of $S^2$ at $\bold n$.
In spherical coordinates $(\theta,\phi)$, for which
$n_1=\sin\theta\cos\phi$, $n_2=\sin\theta\sin\phi$, and $n_3=\cos\theta$, 2-form ${\bold{area}}(S^2)|_{\theta,\phi}=\sin\theta\,\d\theta\wedge\d\phi$.

After collecting the results of the previous subsections together, we arrive at 
$$
s^*\Theta_1^t[3]_{\fii,\bold n}-\tilde k^2(2t)^{1/2}e^{-t[2{\tilde k}^2\Fii^2-(2j+1)\tilde k\Fii +j(j+1)/2+1/8]}\left[\Omega _1(\fii,{\bold n})+\Omega _2(\fii,{\bold n})+{\Cal O}(t^{-1})\right]
$$
where
$$\align
\Omega _1(\fii,{\bold n})&:=\frac{2\tilde k t}{e^{t \tilde k \Fii}}A^{\Fii,{\bold n};t}_{0,0,0}(a,b,c)
C_{0,a} (\fii,{\bold n})\wedge C_{0,b} (\fii,{\bold n})\wedge C_{0,c} (\fii,{\bold n}), \\
\Omega _2(\fii,{\bold n})&:=\frac{2\tilde k t}{e^{t \tilde k \Fii}}\sum_{s=1}^\infty\big[
A^{\Fii,{\bold n};t}_{-s,s,0}(a,b,c)C_{-s,a} (\fii,{\bold n})\wedge C_{s,b} (\fii,{\bold n})\wedge C_{0,c} (\fii,{\bold n})\\
&+A^{\Fii,{\bold n};t}_{0,-s,s}(a,b,c)C_{0,a} (\fii,{\bold n})\wedge C_{-s,b} (\fii,{\bold n})\wedge C_{s,c} (\fii,{\bold n})\big].
\endalign $$
After long but straightforward calculations, one gets
$$\align
\Omega_1(\fii,{\bold n})&=\frac{2 i}{\Fii}d_0 (\fii)\left[\big(e_0 (\fii)+f_0 (\fii)\big)^2+\big(g_0 (\fii)\big)^2\right]\d\fii\wedge{\bold{area}}(S^2)|_{\bold n}\\
&=\frac{i}{\pi}\Fii\left[(1-u)^2+u^2\big(a_0 (\fii)^2+b_0 (\fii)^2\big)+2(1-u)ub_0 (\fii)\right]\d\fii\wedge{\bold{area}}(S^2)|_{\bold n},\\
\Omega _2(\fii,{\bold n})&=4 i d_0 (\fii)\sum_{s=1}^\infty
\frac{\Fii\left[|e_s (\fii)+f_s (\fii)|^2+|g_s (\fii)|^2\right]+
2\,\text{Im}\Big(s\big(\overline{e_s (\fii)}+\overline{f_s (\fii)}\big)g_s (\fii)\Big)}{(\Fii+s)(\Fii-s)}\times\\
&\times
\d\fii\wedge{\bold{area}}(S^2)|_{\bold n}\\
&=\frac{2i}{\pi}u^2\sum_{s=1}^\infty\left[\Fii\left(|a_s (\fii)|^2+|b_s (\fii)|^2\right)
-2 s\,\text{Im}\left(a_s (\fii)b_{-s} (\fii)\right)\right]\d\fii\wedge{\bold{area}}(S^2)|_{\bold n},
\endalign $$
which shows that 
$$
\Omega _1(\fii,{\bold n})+\Omega _2(\fii,{\bold n})=\frac{i}{\pi}F_u(\fii)\,\d\fii\wedge{\bold{area}}(S^2)|_{\bold n}
$$
where
$$\align
F_u(\fii):=&\frac{\fii}{2\pi}\left[(1-u)^2+2(1-u)ub_0 (\fii)\right]\\
&+u^2\sum_{s\in\Z}\left[\frac{\fii}{2\pi}\left(|a_s (\fii)|^2+|b_s (\fii)|^2\right)
-2 s\,\text{Im}\left(a_s (\fii)b_{-s} (\fii)\right)\right].
\endalign $$
To conclude,
$$
s^*\Theta_1^t[3]_{\fii,\bold n}-\tilde k^2(2t)^{1/2}e^{-t[2{\tilde k}^2\Fii^2-(2j+1)\tilde k\Fii +j(j+1)/2+1/8]}\left[
\frac{i}{\pi}F_u(\fii)\,\d\fii\wedge{\bold{area}}(S^2)|_{\bold n}+{\Cal O}(t^{-1})\right] 
\tag{A.1}
$$
Note that if we assume that $\text{supp}\,\alpha\subseteq[1-\epsilon,1]$ where $\epsilon>0$ and $\epsilon\approx 0$, then
$$
F_u(\fii)=\frac{\fii}{2\pi}\left[(1-u)^2+2(1-u)u\frac{\sin\fii}{\fii}+u^2\left(2-\frac{\sin\fii}{\fii}\right)\right]+{\Cal O}(\epsilon).
\tag{A.2}
$$

{\bf The term $\Theta_2^t[3]$.}
Denote $s^*B_2|_{\fii,\bold n}=T_{n,b}S_n^b+T \cdot {\bold 1}$ where $T_{n,b}$ and $T$ are 2-forms
and ${\bold 1}$ is the identity operator.
Then
$$
s^*\Theta_2^t[3]_{\fii,\bold n}=-\sqrt{2t}\tilde k C_{m,a}\wedge T_{n,b}E_{m,n}(a,b)-\sqrt{2t}\tilde k C_{m,a}\wedge T\,
\tr\Big(\psi_m^a e^{-t(h+2i\tilde k \Fii {\bold n\cdot\bold s}+2{\tilde k}^2\Fii^2)}\Big)
$$
where
$$\align
&E_{m,n}(a,b):=\int_{\Delta_{2}}\tr\Big(e^{-t s_1(h+2i\tilde k \Fii {\bold n\cdot\bold s}+2{\tilde k}^2\Fii^2)}\psi_m^a
e^{-t s_2(h+2i\tilde k \Fii {\bold n\cdot\bold s}+2{\tilde k}^2\Fii^2)}S_n^b e^{-t s_3(h+2i\tilde k \Fii {\bold n\cdot\bold s}+2{\tilde k}^2\Fii^2)}\\
&+e^{-t s_1(h+2i\tilde k \Fii {\bold n\cdot\bold s}+2{\tilde k}^2\Fii^2)}S_n^b
e^{-t s_2(h+2i\tilde k \Fii {\bold n\cdot\bold s}+2{\tilde k}^2\Fii^2)}\psi_m^ae^{-t s_3(h+2i\tilde k \Fii {\bold n\cdot\bold s}+2{\tilde k}^2\Fii^2)}\Big)\d s_1\d s_2\\
&=\delta_{n,-m}e^{-t2\tilde k^2\Fii^2}\int_{\Delta_{2}}\Big[
R({\bold n},-t s_2 2\tilde k \Fii)_{a,\tilde a}e^{2\tilde k t s_2 m}
\tr\Big(\psi_m^{\tilde a}S_{-m}^be^{-t 2i\tilde k \Fii {\bold n\cdot\bold s}}e^{-t h}\big)\\
&+R({\bold n},t s_2 2\tilde k \Fii)_{a,\tilde a}e^{-2\tilde k t s_2 m}
\tr\Big(S_{-m}^b\psi_m^{\tilde a}e^{-t 2i\tilde k \Fii {\bold n\cdot\bold s}}e^{-t h}\big)\Big]\d s_1\d s_2.
\endalign $$
Since $S_{-m}^b=T_{-m}^b+K_{-m}^b$, $S_{-m}^b\psi_m^{\tilde a}=\psi_m^{\tilde a}S_{-m}^b-\lambda_{\tilde a b c}\psi_0^c$,
$\tr_b\Big(T_{-m}^be^{-t 2i\tilde k \Fii {\bold n\cdot\bold t}}e^{-t h_b}\Big)\sim\delta_{m,0}2^{-1/2}i j n_b e^{t2\tilde k\Fii j}e^{-t j(j+1)/2}$
we get
$$
E_{m,n}(a,b)\sim-\delta_{m,0}\delta_{n,0}e^{-t[2{\tilde k}^2\Fii^2-(2j+1)\tilde k\Fii +j(j+1)/2+1/8]}\frac{i}{\sqrt{2}}\left(j+\frac{1}{2}\right)n_a n_b.
$$
In addition,
$$\align
&C _{m,a}\tr\left(\psi_m^ae^{-t[h_b+2\tilde k h_f+\frac{1}{8}+2 i\tilde k \Fii({\bold{n\cdot t+n\cdot\underline{k}}})
+2{\tilde k}^2\Fii^2]}\right)\\
&\sim 
-e^{-t[2{\tilde k}^2\Fii^2-(2j+1)\tilde k\Fii+j(j+1)/2+1/8]}\underbrace{n_aC _{0,a}}_{=\d\Fii},
\endalign $$
so that
$$
s^*\Theta_2^t[3]_{\fii,\bold n}\sim
\frac{i}{\pi}\tilde k^2\sqrt{2t}e^{-t[2{\tilde k}^2\Fii^2-(2j+1)\tilde k\Fii+j(j+1)/2+1/8]}\d\fii\wedge\left(\frac{2j+1}{4\tilde k}\frac{n_b T_{0,b}}{\sqrt{2}}+\frac{T}{2 i\tilde k}\right).
\tag{A.3}
$$
Collecting significant terms
$$\align
&s^*\delta\hat\omega_c|_{\fii,\bold n}=\sqrt{2}\big(b_0(\fii)-1\big)S_0^a\epsilon_{abc}\d n_b\wedge\d n_c+\ldots,\\
&s^*\hat\omega^2_c|_{\fii,\bold n}=\frac{1}{\sqrt{2}}S_0^d\epsilon_{d a\tilde a}\big[
\big(a_s(\fii)\d n_a+b_s(\fii)\epsilon_{abc}n_b\d n_c\big)\wedge
\big(a_{-s}(\fii)\d n_{\tilde a}+b_{-s}(\fii)\epsilon_{\tilde a\tilde b\tilde c}n_{\tilde b}\d n_{\tilde c}\big)\\
&+\epsilon_{abc}\epsilon_{\tilde a\tilde b\tilde c}n_bn_{\tilde b}\d n_c\wedge \d n_{\tilde c}
-2\big(a_0(\fii)\d n_a+b_0(\fii)\epsilon_{abc}n_b\d n_c\big)\wedge\epsilon_{\tilde a\tilde b\tilde c}n_{\tilde b}\d n_{\tilde c}\big]\\
&+\tilde k s\big(a_s(\fii)\d n_a+b_s(\fii)\epsilon_{abc}n_b\d n_c\big)\wedge
\big(a_{-s}(\fii)\d n_{a}+b_{-s}(\fii)\epsilon_{a\tilde b\tilde c}n_{\tilde b}\d n_{\tilde c}\big)+\ldots,\\
&s^*\{Q_\bullet,\langle\psi,F\rangle\}_{\fii,\bold n}=2 i \tilde F_{-n}^b S_n^b+2\sqrt{2}\tilde k\Fii n_b\tilde F_0^bI,
\endalign $$
where $\tilde F_{-n}^b$ is a Fourier coefficient of $s^*\langle\psi,F\rangle_{\fii,\bold n}$,
we see that
$$\align
\frac{n_b T_{0,b}}{\sqrt{2}}&=\left\{u^2\left[\sum_{s\in\Z}\left(|a_s(\fii)|^2+|b_s(\fii)|^2\right)+1-2b_0(\fii)\right]
+2u\big(b_0(\fii)-1\big)\right\}{\bold{area}}(S^2)|_{\bold n}\\
&-\frac{u}{2\sqrt{2}}in_b\tilde F_0^b+\ldots,\\
T&=u^2\tilde k\sum_{s\in\Z}(-2 s)\big(a_s(\fii)b_{-s}(\fii)-a_{-s}(\fii)b_s(\fii)\big){\bold{area}}(S^2)|_{\bold n}
-\frac{u}{2}\sqrt{2}\tilde k\Fii n_b\tilde F_0^b+\ldots.
\tag{A.4}
\endalign 
$$

{\bf Conclusions.}
Using the formal notations $\delta(\fii-a)$ for the Dirac measure $\delta_a$
concentrated on $a\in(0,2\pi)$ and $w(\fii)$ for the measure defined by a 1-form $w(\fii)\d\fii$ where $g\in C\big([0,2\pi]\big)$, one gets
$$
\delta(\fii-a)=\mathop{\hbox{w$^*$-lim}}_{r\to\infty}\sqrt{\frac{r}{\pi}}e^{-r(\fii-a)^2}
$$
and for any $p\ge 1$,
$$
\mathop{\hbox{w$^*$-lim}}_{r\to\infty}\frac{1}{r^p}\sqrt{r}\,e^{-r\fii^2}=0.
$$
Since $2{\tilde k}^2\Fii^2-(2j+1)\tilde k\Fii +j(j+1)/2+1/8=\left(\frac{\tilde k}{\sqrt{2}\pi}\right)^2(\fii-\fii_j^k)^2$ where 
$$
\fii_j^k:=2\pi\frac{2j+1}{4\tilde k}=2\pi\frac{2j+1}{k+2}\in(0,2\pi)
$$
and using equatoins (A.1), (A.3), and (A.4),
we may conclude that,
when $s^*\Theta_1^t[3]$ and $s^*\Theta_2^t[3]$ are interpreted as measures on $[0,2\pi]\times S^2$,
$$
\mathop{\hbox{\rm w$^*$-lim}}_{t\to\infty}\left(s^*\Theta_1^t[3]+s^*\Theta_2^t[3]\right)=-\sqrt{\pi}i\left(j+\frac{1}{2}\right)\delta_{\fii_j^k}\otimes{\bold{area}}(S^2).
$$
Proceeding similarly as above, one sees immediately that the weak-star limit of the three-form $s^*\Theta^t_3[3]$
can be interpreted as the zero measure. Thus,
$$
\mathop{\hbox{\rm w$^*$-lim}}_{t\to\infty}s^*\Theta^t[3]=-\sqrt{\pi}i\left(j+\frac{1}{2}\right)\delta_{\fii_j^k}\otimes{\bold{area}}(S^2)
$$
and theorem B follows.

For large $t,$
$$
s^*\Theta^t[1]_{\fii,\bold n}\sim -\sqrt{2t}\tilde k 
e^{-t[2{\tilde k}^2\Fii^2-(2j+1)\tilde k\Fii+j(j+1)/2+1/8]}{\d\Fii}.
$$
Hence, when $s^*\Theta^t[1]$ is interpreted as a measure on $[0,2\pi]$,
$$
\mathop{\hbox{\rm w$^*$-lim}}_{t\to\infty}s^*\Theta^t[1]=-\sqrt{\pi}\delta_{\fii_j^k}.
$$
This proves theorem A.
 


\vskip 0.3in
\bf Appendix: Zeros of $Q_A$ \rm 

\vskip 0.2in

The zero modes of the operator $Q_A$ are the same as its square $Q_A^2$ but the latter is easier to handle.
First we gauge transform the potential $A$ to a constant potential in the Cartan subalgebra $\hm$ and denote by
$\mu\in \hm^*$ its dual with respect to the Killing form. Then 
$$ Q_A^2 = h + \frac{N}{24} + |\tilde k \mu|^2 + 2\tilde k \mu_i h^{i}   \tag{A1}.$$

Acting with $h$ on the highest weight vectors, corresponding to the weight $\lambda +\rho$  in $H_b\otimes H_f,$
gives the eigenvalue  $|\lambda +\rho|^2 -|\rho|^2.$ Here $\rho$ is half the sum of positive roots of $\gm.$ 
In our normalization of the inner product,
$|\rho|^2 = N/24,$ [FdV], and therefore the eigenvalue of  $Q_A^2$ on the highest weight vector is 
$$ Q_A^2 (\lambda) = |\lambda +\rho|^2 + \tilde k^2|\mu|^2 + 2\tilde k (\mu, \lambda +\rho).\tag{A2}$$

When acting on a general vector of weight $\lambda'$ in the representation space the eigenvalue of $h$ is shifted by
$\tilde k(\lambda' -\lambda - \rho)(2d)$ and so the eigenvalue of $Q_A^2$ becomes
$$ Q_A^2(\lambda') =  |\lambda +\rho|^2 + \tilde k^2 |\mu|^2+ 2\tilde k (\mu, \lambda') + 2\tilde (\lambda' -\lambda -\rho)(d).
\tag{A3}$$
We can rewrite this expression as 
$$Q_A^2 (\lambda') = |\lambda +\rho + \tilde k \mu|^2 -+2\tilde k (\lambda' -\lambda -\rho, \mu + \text{\v{$d$}} ), \tag{A4}$$
where \v{$d$} is the dual root of $d.$ 
After action by a Weyl group element of the affine Lie algebra (implemented here by a gauge transformation) we may assume 
that $\mu + \text{\v{$d$}}$ is in the  Weyl  chamber defined by
$$ (\mu + \text{\v{$d$}}, \alpha_i) \leq 0,$$ 
where $\alpha_i$ with $i=0,1,2,\dots \ell$ are the simple roots. 
Since the difference $\lambda +\rho -\lambda'$ is a sum of simple roots, we conclude that the second term in (A4) is nonnegative.
Thus in order that the eigenvalue $Q_A^2(\lambda')=0$ the first term has to vanish. This implies
$$\tilde k \mu = - \lambda -\rho. \tag{A5}$$

From the last equation follows that all the coordinates $(\mu,\alpha_i)$ are strictly negative , by $(\rho,\alpha_i)=1,$ and so 
the second term in (A4) vanishes if and only if $\lambda' -\lambda -\rho = 0.$ We conclude that the only zeros of $Q_A$ are in
the vacuum $\lambda'=\lambda + \rho$ when $\mu +\text{\v{$d$}}$ is in the negative of the fundamental Weyl chamber  and then 
$\mu$ is given by (A5).

\enddocument